\documentclass{ar-1col}
\usepackage[numbers]{natbib}
\usepackage{graphicx}
\usepackage{url}
\usepackage{epsfig,amsmath,amssymb,slashed}
\newcommand{\bra}[1]{\langle #1|}
\newcommand{\ket}[1]{|#1\rangle}

\newcommand{\di}{{\rm d}}
\newcommand{\ii}{i}
\def\ssn{$\sqrt{s}_{\rm NN}$}

\def\spt{{\cal S}}
\def\wT{{\widehat T}}
\def\wj{{\widehat j}}
\def\wJ{{\widehat J}}

\def\wP{{\widehat P}}

\def\wrho{{\widehat{\rho}}}

\newcommand{\tr}{{\rm tr}}  
  
\newcommand{\e}{{\rm e}}

\newcommand{\omegav}{\boldsymbol{\omega}}

\newcommand{\Psibar}{{\overline \Psi}}

\newcommand{\be}{\begin{equation}}
\newcommand{\ee}{\end{equation}}                                                                               
\def\bea{\begin{eqnarray}}
\def\eea{\end{eqnarray}}

\jname{Xxxx. Xxx. Xxx. Xxx.}
\jvol{AA}
\jyear{YYYY}
\doi{10.1146/((please add article doi))}

\begin{document}

\markboth{Becattini and Lisa}{Polarization and Vorticity in the QGP}

\title{Polarization and Vorticity in the Quark Gluon Plasma}

\author{Francesco Becattini$^1$ and Michael A. Lisa,$^2$
\affil{$^1$Dipartimento di Fisica e Astronomia, University of Florence, Florence, Italy, I-50019; email: becattini@fi.infn.it}
\affil{$^2$Department of Physics, The Ohio State University, Columbus, Ohio, USA 43210; email: lisa.1@osu.edu}}

\begin{abstract}
The quark-gluon plasma produced by collisions between ultra-relativistic heavy nuclei	
  is well described in the language of hydrodynamics.
Non-central collisions are characterized by very large angular momentum,
  which	in a fluid system manifests as flow vorticity.
This rotational structure can lead to a spin polarization of the hadrons that eventually
  emerge from the plasma, providing experimental access	to flow	substructure at
  unprecedented	detail.
Recently, first observations of $\Lambda$ hyperon polarization along the direction
  of collisional angular momentum have been reported.
These measurements are in broad	agreement with hydrodynamic and	transport-based
  calculations and reveal that the QGP is the most vortical fluid ever observed.
However, there remain important tensions between theory and observation	which might be
  fundamental in nature.
In the relatively mature field of heavy ion physics, the discovery of
  global hyperon polarization and three-dimensional simulations	of the collision have
  opened an entirely new direction of research.
We discuss the current status of this rapidly developing area and directions for future	research.
\end{abstract}

\maketitle

\tableofcontents

\vspace*{-2mm} 
\section{INTRODUCTION}
Collisions between heavy nuclei at ultra-relativistic energies create a Quark-Gluon Plasma (QGP)~\cite{Shuryak:1980tp,Adams:2005dq,Adcox:2004mh,Back:2004je,Arsene:2004fa}, characterized
by colored partons as dynamic degrees of freedom.
For more than two decades, a large community has systematically studied
  these collisions to extract insight about quantum chromodynamics (QCD) matter
  under extreme conditions. 
The resulting field of relativistic heavy ion physics is by now
  relatively mature. 
With the early realization that the QGP in these collisions
  is a "nearly perfect fluid,” hydrodynamics has been the dominant theoretical
  framework in which to study the system.

Much of the evidence for the fluid nature of the QGP has been based on the response of the
  bulk medium to azimuthal (to the beam direction) anisotropies in the initial
  energy density~\cite{Heinz:2013th}.
Measured azimuthal correlations are well reproduced by modulations in the
  outward-directed flow fields in the hydro simulations.
However, despite the fact that heavy ion collisions involve huge angular momentum densities
  ($10^{3-4}\hbar$ over volumes $\sim250~{\rm fm}^3$), relatively less focus has
  been placed on the consequences of this angular momentum.

In any fluid, angular momentum manifests as  vorticity 
  in the flow field. The coupling between rotational motion and quantum spin can lead, in the QGP, to polarization of
  hadrons emitted from fluid cells, driven by the local vorticity of the cell.
In 2017, the first experimental observation of vorticity-driven polarization
  in heavy ions was reported \cite{STAR:2017ckg}.
This has generated an intense theoretical activity and further experimental study.
This manuscript
  reviews the tremendous progress and current understanding of the vortical nature of the QGP.
This line of investigation, only just now begun, represents one of the few truly “new” directions in
  the soft sector of relativistic heavy ion physics for many years.

In the next section, we place these studies into a larger context of similar phenomena in
  other physical systems and define geometrical conventions required for the heavy ion case.
We then discuss theoretical tools employed to model the complex rotational
  dynamics of the plasma and the manifestation in particle polarization.
In section~\ref{sec:Observations}, we discuss experimental measurements and observational
  systematics.
We will see broad agreement between observation and theory, but tension in some important
  aspects.
We conclude our review with open questions and an outlook.

\begin{figure}[t]
\includegraphics[width=\textwidth]{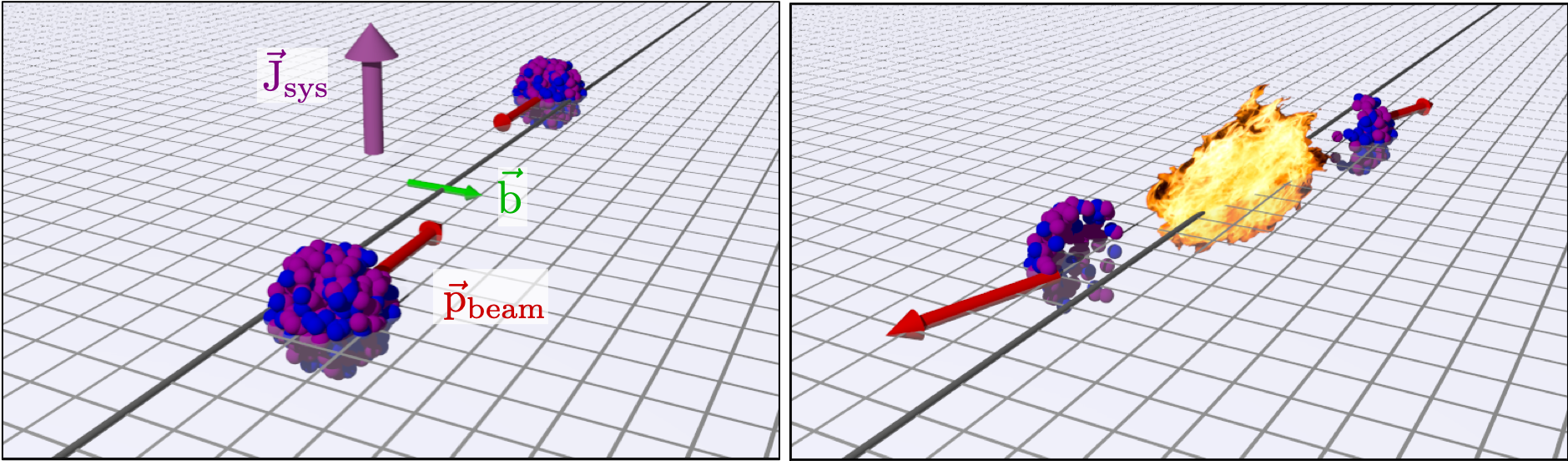}
\caption{
A heavy ion collision at relativistic energy is sketched, in the center of mass frame. 
The relevant geometrical and physical quantities characterizing a collisions are shown in 
the left panel. The Quark Gluon Plasma is formed out of the colliding nucleons of the nuclear overlapping region (right panel). The spectator deflection in the right panel is greatly exaggerated 
for clarity.
}
\label{fig:CollisionSketch}
\end{figure}

\section{QUARK GLUON PLASMA, HYDRODYNAMICS AND VORTICITY}
\label{qgp}

That the QGP produced in collisions of nuclei at relativistic energies is, for a transient 
of around $10^{-22}$ seconds, a nearly perfect fluid is based on the accumulated evidence collected 
over a time span of more than ten years. The main fact is that this fluid breaks up into hadrons 
in a state very close to local thermodynamic equilibrium~\cite{Becattini:2009fv} at a 
temperature very close to the pseudo-critical QCD temperature of 160 MeV \cite{Aoki:2006we},
\cite{Aoki:2006br}.

\begin{figure}[t]
  \begin{minipage}[b]{0.5\textwidth}
    \includegraphics[width=\textwidth]{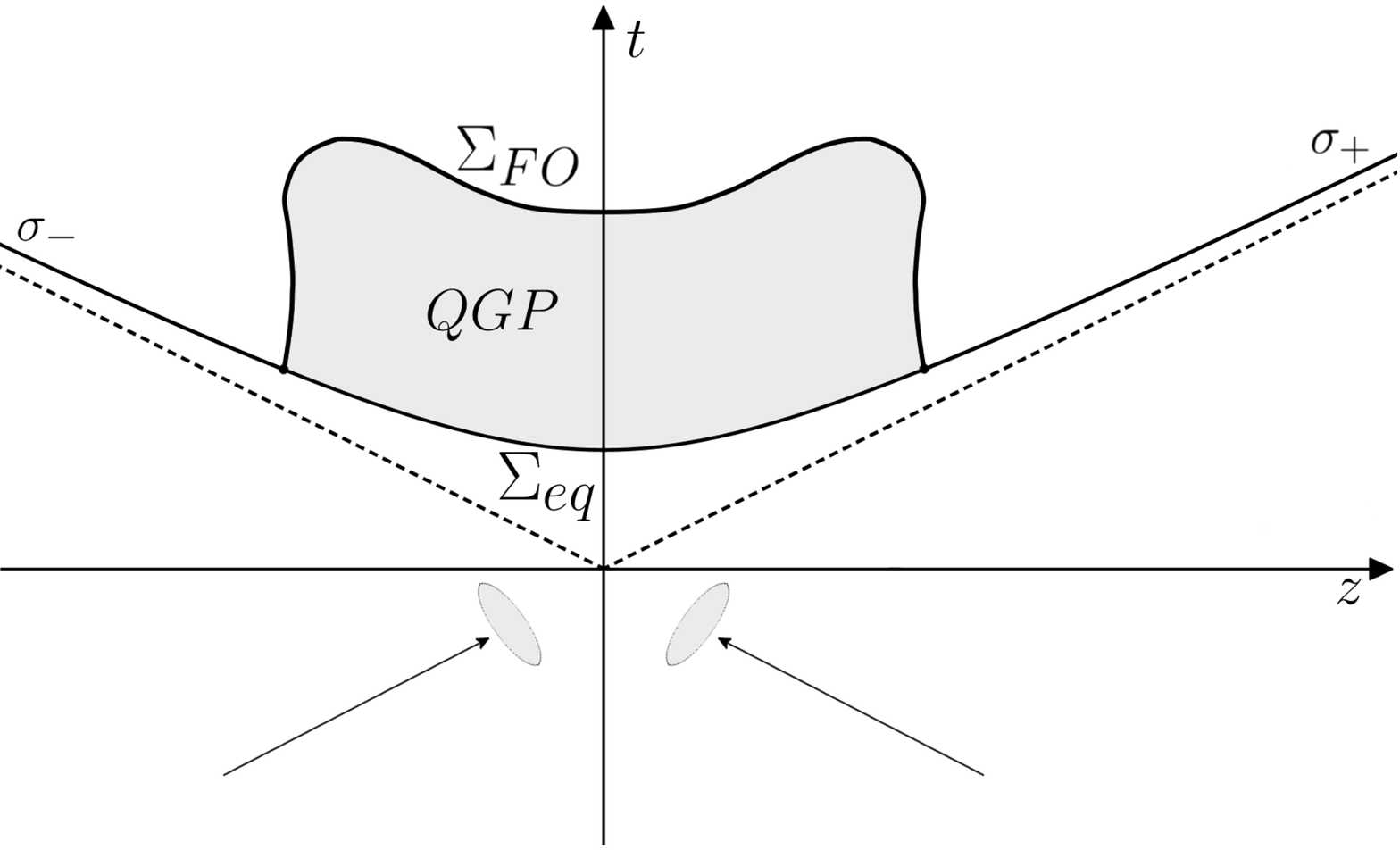}
  \end{minipage}
  \hfill
  \begin{minipage}[b]{0.5\textwidth}
    \includegraphics[width=\textwidth]{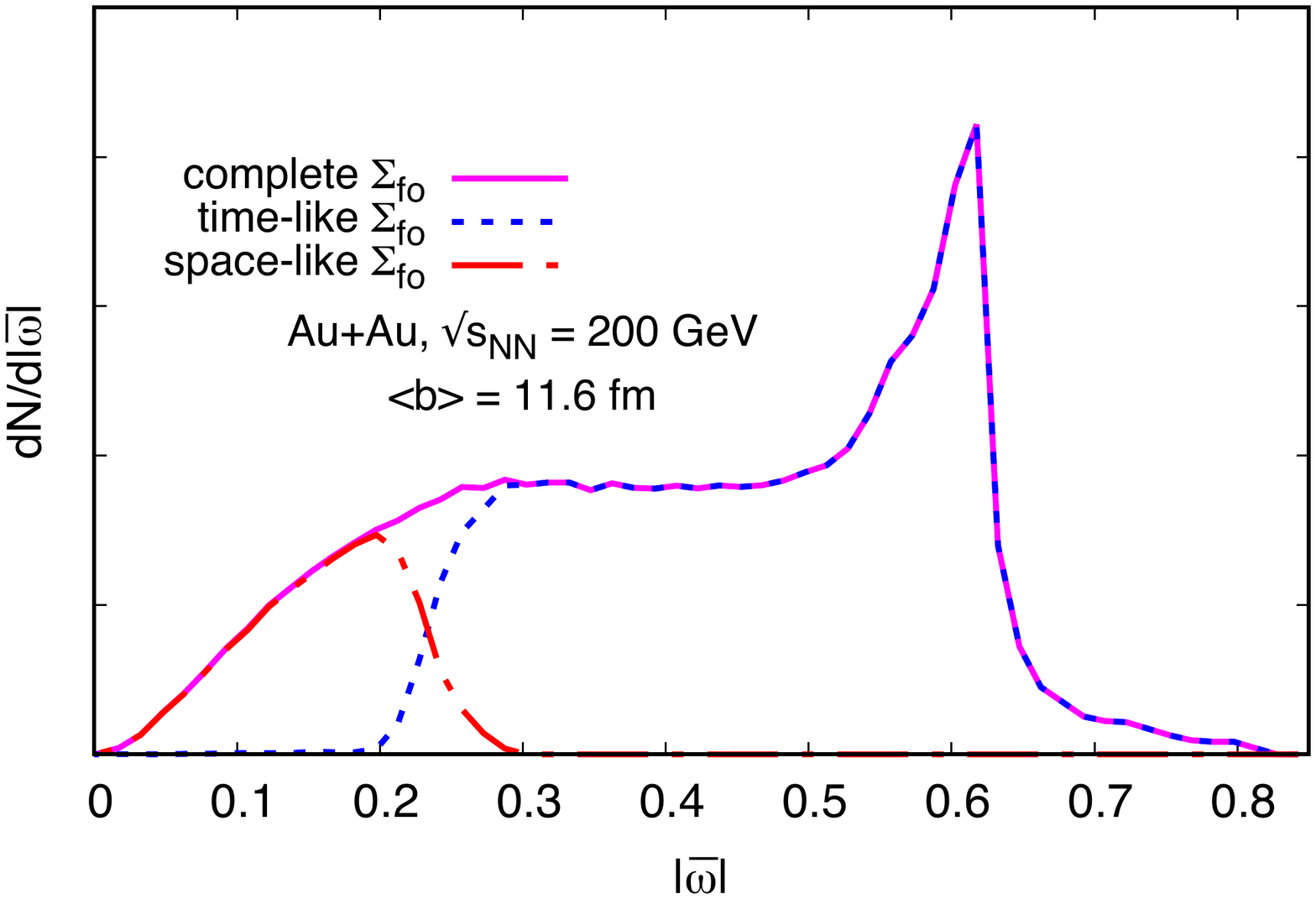}
  \end{minipage}
\caption{Left: \label{freezeout} A collision of two nuclei in space-time diagram. In the hydrodynamic model,
local equilibrium is believed to occur on the hyperbola $\Sigma_{eq}$ which is the 
initial 3D hypersurface of the thermalized QGP and to cease at the 3D hypersurface
$\Sigma_{FO}$\newline
Right: \label{vorthist}
 Distribution of the amplitude of thermal vorticity $|\varpi| = |\sqrt{|\varpi_{\mu\nu} 
\varpi^{\mu\nu}|}$
at the freeze-out hypersurface $\Sigma_{FO}$, calculated with the ECHO-QGP code under
the same conditions as in \cite{Becattini:2015ska} with a freeze-out temperature of 130 MeV. 
The red dashed line is the contribution of the space-like part while the blue dashed line that of the 
time-like part.}
\end{figure}

Local thermodynamic equilibrium implies that momentum spectra of produced hadrons are
very well reproduced by the assumption of a local Bose-Einstein or Fermi-Dirac distribution
function (for vanishing chemical potentials):
\be\label{phsdf}
   f(x,p) = \frac{1}{\exp[\beta \cdot p] \pm 1}
\ee
where $\beta = (1/T)u(x)$ is the {\em four-temperature} vector including temperature and the
four-velocity hydrodynamic field $u(x)$. The formula~\ref{phsdf} applies to the local fluid
cell, and should be integrated thereafter over the ``freeze-out" 3D-hypersurface 
(see figure~\ref{freezeout}) defined as the boundary of local thermodynamic equilibrium,
giving rise to what is well known in the field as ``Cooper-Frye" formula~\cite{Cooper:1974mv}.
Indeed, this is analogous to the last-scattering surface in the cosmological expansion where 
the background electro-magnetic radiation froze out. 

Apparently, local thermodynamic equilibrium is achieved - and a plasma at finite temperature
is formed - at quite an early time in the process (see figure~\ref{freezeout}). This is confirmed
by the success of the hydrodynamic equations in determining the flow field $u(x)$ in 
eq.~\ref{phsdf}. Particularly, the model is able to successfully account for the observed
anisotropies of the momentum spectra in the transverse plane perpendicular to the beam line 
(refer to fig.~\ref{fig:CollisionSketch}) as a function of the azimuthal angle. These 
anisotropies - encoded in the Fourier coefficients $v_n$ - have led to the conclusion that the 
viscosity of the QGP must be very small as compared to the entropy density, close to 
the conjectured universal lower bound of $\hbar/4\pi$ \cite{Kovtun:2004de}. 

Recently, the exploration of the QGP made a significant advance. The measurement of polarization 
of emitted hadrons made it clear that a new probe is accessible which may give a wealth of 
new and complementary information. In particular, in the hydrodynamic paradigm, while the 
momentum spectra provide direct information about the velocity and the temperature field, 
polarization is linked to the vorticity and more generally to the {\em gradients} of these 
fields (see Section~\ref{theory}). 
This is an interesting aspect. In ideal hydrodynamics, particle distribution function, such as 
eq.~\eqref{phsdf} is determined by intensive thermodynamic quantities of the local cell in 
the local rest frame, such as temperature and chemical potentials related to the various charges 
(baryon number, electric charge, etc). Likewise, assuming that spin degrees of freedom locally 
equilibrate, vorticity plays the role of a potential determining the "spin charge" distribution
of particles, e.g. number of spin up versus spin down (see Section \ref{theory}). Vorticity 
should be then considered a further intensive thermodynamic quantity needed to describe locally 
the fluid. In a sense, vorticity is an extra substructure of a hydrodynamic cell. This
property makes polarization a very sensitive probe of the dynamical process leading to the QGP 
formation and of its evolution. As has been mentioned in the Introduction, this field has only 
just begun and all the developments that polarization may lead to can be hardly envisioned 
for the present.  

\subsection{Vorticity and polarization: overview}
\label{sec:rotpol}

While the QGP formed in heavy ion collisions is only a few times larger than a
  nucleus, "heavy" ions are utilized in order to form a bulk system, significantly
  larger than the confinement volume characteristic of a hadron. Otherwise stated, 
  the system which is formed is much larger than the typical microscopic interaction
  scale, and such a separation of scales (``hydrodynamic limit") makes it possible 
  to talk about a fluid and to use hydrodynamics as an effective tool to describe 
  its evolution; in the hydro language, the Knudsen number is sufficiently small 
  \cite{Niemi:2014wta}. 
Under such circumstances, the variation of the flow field in space and time can
be slow enough to be dealt with as "macroscopic" motion of bulk matter and vorticity
as well. As it will become clear later on in Section~\ref{theory}, the vortical 
structure is {\it probed} by the spin of hadrons that "freeze out" from local fluid 
cells in a state of local thermodynamic equilibrium, as has been discussed above.
More specifically, the presence of a vortical motion (as well as an acceleration and
a temperature gradient) entails a modification of~\ref{phsdf} such 
that the distribution function becomes non-trivially dependent on the spin degrees
of freedom. 
  
That spin and vorticity are tightly related is not a new insight, and yet there 
are relatively few examples of physical systems which show the effect of the coupling 
between mechanical angular momentum of bulk matter and the quantum spin of particles 
that comprise (or emerge from) that matter.

Two seminal measurements were reported nearly simultaneously more than a century ago.
Barnett~\cite{PhysRev.6.239} observed that an initially-unmagnetized steel cylinder
 would generate a magnetic field upon being rotated.
In the same year, Einstein and de Haas~\cite{1915KNAB...18..696E} observed the complementary
  effect: a stationary unmagnetized ferromagnetic object will begin to rotate upon
  introduction of an external magnetic field.
In both cases, the phenomenon is rooted in the conservation of total angular
 momentum on one hand and on equipartition of angular momentum, that is thermodynamic 
 equilibrium, on the other. In the Barnett
 effect, the angular momentum which is imparted through a forced rotation gets partly
 distributed to the quantum spin of the constituents and, once thermodynamic equilibrium
 is reached, a stable magnetic field is generated as a consequence of the polarization
 of matter. In the Einstein-de Haas effect, the external magnetic field implies, at
 thermodynamic equilibrium, a polarization of matter, whence an angular momentum; if
 the magnetic field does not provide torque, the body should start spinning as to 
 conserve the initial vanishing angular momentum. 
Indeed, a quantitative understanding of these phenomena was possible only a decade
  later, with the discovery of the electron spin and anomalous gyromagnetic ratio.

Another example is found in low-energy heavy ion reactions, in which a beam with
  kinetic energy of $E_{\rm kin}\sim30$~MeV per nucleon is incident on a stationary target.
(In high-energy physics terms, $\sqrt{s_{NN}}-2m_p\approx15$~MeV, where $m_p$
  is the proton mass.)
This is the regime of quasi compound nucleus formation, in which the short-lived
  system is assumed to rotate as a whole, to first order.
At "high" beam energies $E_{\rm kin}\gtrsim50$~AMeV, projectile fragments are expected
  to experience positive deflection (c.f. section~\ref{sec:Conventions}) due
  to collisional and bulk compression during the collision.
At lower energies, collisions are Pauli-suppressed and attractive nuclear surface
  interactions are expected to produce an orbiting motion that leads to negative deflection.
Disentangling the interplay between these physical mechanisms requires determination
  of $\hat{J}_{\rm sys}$.
This was achieved by correlating~\cite{PhysRevLett.57.559,Lemmon:1999lnn}
  the circular polarization of $\gamma$ rays with forward fragment deflection angles.
These measurements represent the first observation of "global polarization" in (nonrelativistic)
  heavy ion reactions.

In the above cases, the bulk mechanical motion is basically rigid-body rotation.
\footnote{Low-energy compound nuclei have surface vibrations and breathing, but generally
  do not feature internal fluid flow structure.}
Only recently~\cite{TakahashiFluidSpintronics} has mechanically-induced spin polarization
  been observed in a fluid.
Liquid Hg flowing through a channel experiences viscous forces along the channel walls,
  generating a local vorticity field whose strength and direction varies as a function
  of position.
Hydrodynamic vorticity-spin coupling then produces a corresponding electron polarization
  field, which was measured using the inverse spin Hall effect~\cite{InverseSpinHallEffect}.
This experiment, where both the vorticity and the induced polarization are observable,
  is important to establish the phenomenon in fluids.

With respect to all above listed cases, polarization in relativistic heavy ions possesses
two unique features. First of all, its measurement is not mediated by a magnetic 
field (like in the Barnett effect) but the mean spin of particles is directly observed; 
this is not possible in ordinary matter.
Secondly, and maybe more importantly, the system at hand - QGP at very high energy - is almost 
neutral by charge conjugation, i.e. C-even. If it was precisely neutral, the observation
of polarization by magnetization would be simply impossible because particles and antiparticles 
have opposite magnetic moment. In fact, as we will see, $\Lambda$ and $\bar\Lambda$ in
relativistic nuclear collisions at high energy have
almost the same mean spin, which is an unmistakable signature of a thermal-mechanical
driven polarization. If the electro-magnetic, or any other C-odd mean field, was responsible
for this effect, the sign of the mean spin vector components would be opposite. 
Hence, altogether, while for non-relativistic matter (without anti-matter) it is impossible to 
resolve polarization by rotation and by magnetization - what lies at the very heart of
the Barnett and Einstein-de Haas effects - in relativistic matter, because of the 
existence of antiparticles, they can be distinguished 
and QGP is the first relativistic system where the distinction has been observed.

\subsection{Geometry of a nuclear collision}
\label{sec:Conventions}

The left panel of figure~\ref{fig:CollisionSketch} sketches the geometry of a heavy ion collision in
  its center of mass frame, prior to contact.
Designating one nucleus the beam and the other the
  target\footnote{This initial designation is of course arbitrary, but the convention
  must be kept consistently.  In the age of collider-mode nuclear
  physics, confusion is not uncommon and leads to sign errors.}, the impact parameter $\vec{b}$ points from
  the center of the target to the center of the beam, perpendicular to the beam momentum $\vec{p}_{\rm beam}$.
The vectors $\vec{b}$ and $\vec{p}_{\rm beam}$ span the reaction plane, indicated by the grid.
The total angular momentum of the collision $\vec{J}_{\rm sys}=\vec{b}\times\vec{p}_{\rm beam}$.

The right panel sketches the situation after the collision.
In the participant-spectator model~\cite{Westfall:1976fu} commonly used at high energies,
  a fireball at midrapidity is produced by the sudden and violent deposition of energy when
  "participant" nucleons overlap and collide.
Meanwhile, projectile nucleons that do not overlap with oncoming nucleons in the target
  are considered "spectators" and continue with their forward motion essentially unchanged,
  later to undergo nuclear fragmentation.

However, this distinction is not so sharp in reality, as the forward "spectators" do receive a
  sideways repulsive impulse during the collision, as indicated by the deflected momentum arrows
  in the right panel of figure~\ref{fig:CollisionSketch}.
The case shown in the figure is deflection to "positive" angles, to distinguish the case
  at much lower energy~\citep[e.g.][]{Lemmon:1999lnn} where attractive forces can
  produce negative deflection.
In the parlance of relativistic collisions~\cite{Poskanzer:1998yz}, the positive deflection corresponds to positive
  directed flow ($v_1$) in the forward direction ($v_1 > 0$ when $y\approx y_{\rm beam}$).

This deflection is important.
While we are especially interested in the vortical structure of the fireball at midrapidity,
  we need to know the direction of the angular momentum, which must by symmetry give the
  average direction of vorticity.
Forward detectors are used to estimate $\hat{J}_{\rm sys}$ event-by-event, as discussed below.

A final note about coordinate systems and conventions.
It is common to define a coordinate system in which $\hat{z}\equiv\hat{p}_{\rm beam}$ and
  $\hat{x}\equiv\hat{b}$.
In this case, $\hat{J}=-\hat{y}$; 
The azimuthal angle of $\hat{b}$ about $\hat{p}_{\rm beam}$ in some coordinate system
  (say, the floor of the experimental facility)
  is often referred to as the reaction plane angle $\Psi_{RP}$.
The aforementioned forward detectors use spectator fragment deflection to  
  determine the event plane angle $\Psi_{\rm EP,1}$.
Standard techniques have been developed~\cite{Poskanzer:1998yz} to determine
  the event plane and the resolution with which it approximates $\Psi_{RP}$,
  i.e. the direction $\hat{b}$.

Since the size and angular momentum of the QGP fireball depends on the overlap between
  the colliding nuclei, an estimate of the magnitude of the impact parameter is also important.
Standard estimators~\cite{Miller:2007ri}, typically based on the charged particle multiplicity measured at midrapidity,
  quantify the "centrality" of each collision in terms of fraction of inelastic cross-section.
Head-on ($|\vec{b}|=0$) and barely-glancing collisions are said to have centrality of 0\% and 100\% respectively.

\section{POLARIZATION IN RELATIVISTIC HEAVY ION COLLISIONS: THEORY} 
\label{theory}

The main purpose of the theoretical work is to calculate the amount of polarization of observable
particles once the initial condition of the collision is known, that is the energy and the 
impact parameter of the two nuclei. The final outcome depends on the model of the collision 
(see Section~\ref{qgp})
and on how the initial angular momentum may induce a global polarization of the particles. 

The first calculation on global polarization in relativistic heavy ion collision was presented
in ref.~\cite{Liang:2004ph} based on a perturbative-QCD inspired model where colliding partons
get polarized by means of a spin-orbit coupling. The amount of predicted polarization of 
$\Lambda$ baryons was originally large (around 30\%) and corrected thereafter by the same 
authors to be less than 4\% \cite{Gao:2007bc}. Besides the apparent large uncertainty, the main 
problem of the collisional approach - at the quark-gluon level - is the difficulty of reconciling 
it with the evidence of a strongly interacting QGP, which makes the kinetic approach dubious. 
Another problem is how to transfer the polarization at quark-gluon level to final hadrons, 
which requires a detailed hadronization model and more assumptions. This scenario, however, 
has been further developed and it will be addressed later in this Section. 

About the time when the first measurement of global $\Lambda$ polarization at RHIC appeared 
\cite{Abelev:2007zk} setting an upper limit of few percent, the idea of a polarization related 
to hydrodynamic motion, and particularly vorticity, was put forward \cite{Abreu:2007kv,Betz:2007kg}.
If the QGP achieves and maintain local 
thermodynamic equilibrium until it decouples into freely streaming 
non-interacting hadrons and if this model - as discussed in the Introduction and
in Section \ref{qgp} - is very successful to describe the momentum spectra 
of particles, there is no apparent reason why it should not be applicable to the spin degrees 
of freedom as well. Hence, polarization must be derivable from the very fact that the system is 
at local thermodynamic equilibrium, whether in the plasma phase or in the hadron phase just before 
they freeze-out. This idea establishes a link between spin and vorticity (more precisely thermal 
vorticity as later described) and makes it possible to obtain quantitative predictions at hadronic
level without the need of a mechanism to transfer polarization from partons to hadrons. The 
actual quantitative relation for a {\em relativistic} fluid was first worked out in global 
equilibrium \cite{Becattini:2007nd}, then at local equilibrium for spin 1/2 particles in 
ref.~\cite{Becattini:2013fla}.

For a particle with spin 1/2 the mean spin vector is all is needed to describe polarization 
(this is not the case for spin greater than 1/2) and the relativistic formula was found to be, 
at the leading order \cite{Becattini:2013fla}:
\be\label{basic}
 S^\mu(p)= - \frac{1}{8m} \epsilon^{\mu\rho\sigma\tau} p_\tau 
 \frac{\int d\Sigma_\lambda p^\lambda n_F (1 -n_F) \varpi_{\rho\sigma}}
  {\int d\Sigma_\lambda p^\lambda n_F}
\ee
where $p$ is the four-momentum of the particle, and $n_F = (1+\exp[\beta \cdot p - \mu Q/T] +1)^{-1}$ 
is the Fermi-Dirac distribution with four-temperature $\beta$ like in eq.~\ref{phsdf} and with 
chemical potential $\mu$ coupled to a generic charge $Q$. The integration should be carried out 
over the freeze-out hypersurface (see fig.~\ref{freezeout}); in a sense, in the heavy ion jargon 
this can be called the "Cooper-Frye" formula for the spin. The key ingredient in  equation~\ref{basic} 
is the so-called {\em thermal vorticity} tensor $\varpi(x)$, which reads: 
\be\label{thvort}
   \varpi_{\mu\nu} = -\frac{1}{2} \left( \partial_\mu \beta_\nu - \partial_\nu 
   \beta_\mu \right)
\ee
i.e. the anti-symmetric derivative of the four-temperature. This quantity is adimensional
in natural units and it is the proper extension of the angular velocity over temperature 
ratio mentioned in the Introduction. Hence the spin depends, at the leading order, 
on the {\em gradients} of the temperature-velocity 
fields, unlike momentum spectra which depend, at the leading order, on the temperature-velocity
field itself. Thereby, polarization can provide a complementary information about the
hydrodynamic flow with respect to the spectra and their anisotropies. The formula \ref{basic}
applies to anti-particles as well, so that in a charge-neutral fluid the spin vector is
expected to be the same for particles and anti-particles, which is a remarkable feature
as emphasized in Subsection~\ref{sec:rotpol}.
It is worth pointing out that  formula~\ref{basic} implies that a particle within a fluid 
in motionvat some space-time point $x$ gets polarized according to (natural constants have been 
purposely restored):
\be\label{restframe}
{\bf S}^*(x,p) \propto \frac{\hbar}{KT^2} \gamma {\bf v} \times \nabla T + 
\frac{\hbar}{KT} \gamma (\omegav - (\omegav \cdot {\bf v}) {\bf v}/c^2) 
+ \frac{\hbar}{KT} \gamma {\bf A} \times {\bf v}/c^2
\ee
where $\gamma = 1/\sqrt{1-v^2/c^2}$ and all three-vectors, including vorticity, acceleration 
and velocity, are observed in the particle rest frame. The decomposition~\ref{restframe} 
makes it clear what are the thermodynamic "forces" responsible for polarization: the last 
term corresponds to the acceleration-driven polarization, its expression is reminiscent of 
the Thomas precession and it is indeed tightly related to it (particle moving in an accelerated 
flow); the second term is the relativistic expression of polarization by vorticity; the 
first term is a polarization by combination of temperature gradient and hydrodynamic flow 
and is, to the best of our knowledge, a newly found effect.

First hydrodynamic calculations based on formula~\ref{basic} predicted a global 
polarization of $\Lambda$ baryons of a few percent at \ssn = 200 GeV \cite{Becattini:2013vja}, 
hence compatible with the previous experimental limit. The new measurements with a larger 
statistics then confirmed that polarization value is of such order of magnitude.  
Formula~\ref{basic} then became a benchmark for most phenomenological calculations of
the polarization in heavy ion collisions. 

We will now review in some detail the status of the theoretical understanding of the
polarization in relativistic fluids and in nuclear collisions particularly.

\subsection{Polarization in statistical mechanics}
\label{sec:PolarizationStatMech}

The calculation of spin at global or local thermodynamic equilibrium requires a quantum framework, 
spin being inherently a quantum observable. The most appropriate framework is thus quantum 
statistical mechanics, and since we are dealing with a relativistic fluid, in a relativistic 
setting. However, many quantitative features can be found out starting from the simplest 
non-relativistic case.

As a simple illustrative case, consider a rotating ideal gas with angular velocity $\omegav$ 
within a cylindrical vessel of radius $R$. At equilibrium, the statistical operator 
\footnote{We will denote quantum operators with an upper wide hat throughout.} reads 
\cite{Landau:1980mil}:
\be\label{densop1}
\wrho = \frac{1}{Z} \exp \left[ -\frac{\widehat H}{T} + 
 \frac{\omegav \cdot \widehat{\bf J}}{T} \right]
\ee
Since the particles are free, both the Hamiltonian and angular momentum operator are the sum of
individual single-particle operators and the density operator can be factorized. Since the
total angular momentum includes both orbital and spin part, that is $\widehat{\bf J}_i = 
\widehat{\bf L}_i + \widehat{\bf S}_i$ for each particle $i$, the spin density matrix for 
a particle with momentum $p$ turns out to be:
\be\label{spindensmat}
  \Theta(p)_{rs} \equiv \bra{p,s} \wrho_i \ket{p,r} = \frac{ \bra{p,s} 
   \exp[\omegav \cdot \widehat{\bf S}_i/T] \ket{p,r} }{\sum_{t=-S}^S 
    \bra{p,t} \exp[\omegav \cdot \widehat{\bf S}_i/T] \ket{p,t} } = 
    \delta_{rs} \frac{ \exp[ - s \omega/T]} {\sum_{t=-S}^S \exp[- t \omega/T]}
\ee
implying a mean spin vector of particles:
\be\label{meanspin1}
 {\bf S} = \hat\omegav \frac{\partial}{\partial (\omega/T)} 
 \frac{\sinh[(S+1/2)\omega/T]}{\sinh[\omega/2T]} \simeq \frac{S(S+1)}{3} \frac{\omegav}{T}
\ee
where the last expression is the leading order for small ratios $\omega/T$. 
Equation~\ref{spindensmat} also implies that the so-called alignment $\Theta_{00}$ for 
spin 1 particles is quadratic in $\omega/T$ at the leading order, which puts a severe 
limitation to its observability in relativistic heavy ion collisions (see also subsection \ref{sec:align}).

In the more general, relativistic case, the equilibrium operator~\ref{densop1} is replaced by 
\cite{Becattini:2012tc}:
\be\label{densop2}
\wrho = \frac{1}{Z} \exp \left[ - b_\mu \wP^\mu + \frac{1}{2} 
 \varpi_{\mu\nu} \wJ^{\mu\nu} \right]
\ee
where $b$ is a constant time-like four-vector and $\varpi$ is the thermal vorticity which, 
at global thermodynamic equilibrium ought to be constant; $\wP$ and $\wJ$ are the four-momentum
and angular momentum-boost operators. It is important to point out that 
thermal vorticity includes both vorticity {\em and} acceleration besides the gradient of 
the temperature. For instance, at global equilibrium, it turns out \cite{Becattini:2015nva}:
\be
 \varpi_{\mu\nu} = \frac{1}{2} \varepsilon_{\mu\nu\rho\sigma} \frac{1}{T} 
  \omega^\rho u^\sigma + \frac{1}{T} \left( A_\mu u_\nu - A_\nu u_\mu \right)
\ee
where $u$ is the four-velocity, $A$ the four-acceleration and $\omega$ the vorticity
four-vector. The entanglement of vorticity and acceleration is a typical signature of 
relativity, much like that of electric and magnetic field in the electromagnetic
tensor $F_{\mu\nu}$.

An intermediate step towards formula~\ref{basic} is the free single-particle quantum
relativistic calculation. In this case, for a single particle, the operator~\ref{densop2} 
leads to the spin density matrix \cite{Becattini:2019ntv}:
\be\label{spindensmat2}
  \Theta(p) = \frac{D^S([p]^{-1} \exp[(1/2) \varpi : \Sigma_S] [p])+ D^S([p]^{\dagger} 
   \exp[(1/2) \varpi : \Sigma^\dagger_S] [p]^{-1\dagger})}
   {\tr (\exp[(1/2) \varpi : \Sigma_S] + \exp[(1/2) \varpi : \Sigma_S^\dagger])},
\ee
where $D^S( )$ stands for the $(2S+1)$-dimensional representation of the group SL(2,C)
universal covering of the Lorentz group, $\Sigma_S$ are the $(2S+1) \times (2S+1)$ matrices 
representing the Lorentz generators, $[p]$ is the so-called {\em standard Lorentz 
transformation} which takes the unit time vector $\hat t$ into the direction of the
four-momentum $p$ \cite{WuKiTung:1985}. The spin density matrix in eq.~\ref{spindensmat2} 
implies a mean spin four-vector, for sufficiently low values of the thermal vorticity:
\be\label{meanspin2}
 S^\mu(p) = - \frac{1}{2m}\frac{S(S+1)}{3} \epsilon^{\mu\alpha\beta\nu} 
  \varpi_{\alpha\beta} p_\nu,
\ee
which is a direct relativistic extension of the formula~\ref{meanspin1} \cite{Becattini:2016gvu}.

For a system of many particles, just like those emerging from a nuclear collisions, one
would take the formula~\ref{meanspin2} and average it over the different
particle-emitting spots, i.e. over the hydrodynamic cells of the freeze-out hypersurface. 
The result is the formula~\ref{basic} except the factor $(1-n_F)$. Indeed, the latter is the 
typical signature of Fermi statistics and it naturally comes out in a quantum-field
theoretical calculation. Indeed, this was the approach taken into the original calculation
at {\em local} thermodynamic equilibrium in ref.~\cite{Becattini:2013fla} where the 
density operator is the extension of equation~\ref{densop2}:
\be\label{ledo} 
  \wrho = \dfrac{1}{Z} \exp \left[ -\int_{\Sigma_{FO}} \di \Sigma \; n_\mu 
   \left( \wT^{\mu\nu}(x) \beta_\nu(x) - \zeta(x) \wj^\mu(x) \right) \right]
\ee  
where $\beta_\nu(x)$ is the four-temperature function (dependent on space and time),
$\zeta(x)$ is the ratio between chemical potential and temperature and $\wT$,$\wj$
are the stress-energy tensor and current operators respectively. The integration 
should be done over the freeze-out 3D-hypersurface (see figure~\ref{freezeout}) which
is supposedly the boundary of local thermodynamic equilibrium. Indeed, calculating
the mean spin vector from the density operator~\ref{ledo} is not straightforward
and some key assumptions are needed to get to the formula~\ref{basic}. The most 
important is the usual hydrodynamic limit: microscopic lengths should be much smaller 
than the hydrodynamic scale, that is $\beta(x)$ should be a slowly varying function.  
The second main assumption used in the original calculation \cite{Becattini:2013fla} 
was an {\em ansatz} for the covariant Wigner function at global equilibrium with 
acceleration and rotation, that is with the density operator~\ref{densop2}. In spite 
of these assumptions, there are good reasons to believe that the exact formula at 
the leading order in thermal vorticity in a quantum field theory calculation would 
precisely be equation.~\ref{basic}. Indeed, the same formula was found with a different
approach, based on the $\hbar$ expansion of the Wigner equation \cite{Fang:2016vpj}
and, furthermore, it is the only possible linear expression in $\varpi$ yielding the 
correct single-particle \eqref{meanspin2} and non-relativistic limit. For instance,
a term proportional to $\varpi^{\mu\nu} p_\nu$, even if orthogonal to $p$, would
not yield correct limiting cases. What is still unknown is the exact global equilibrium 
formula at all orders in thermal vorticity including quantum statistics.

While the local equilibrium calculation of the spin density matrix and related quantities
at leading order seems to be established at the most fundamental level of quantum
field theory, some questions remain to be addressed. It is not known how large are the 
higher order terms in thermal vorticity at local equilibrium, nor we have an exact 
solution at global equilibrium with the density operator~\ref{densop2} including 
quantum field effects, namely quantum statistics. Very little is known about the 
the dissipative, non local-equilibrium terms, and their magnitude. Recently, a phenomenological 
approach to spin dissipation has been taken \cite{Hattori:2019lfp} generalizing a familiar 
classical method to constrain constitutive equations in dissipative hydrodynamics, based on the
positivity of entropy current divergence \cite{Israel:1976tn}. It remains to be understood 
whether such a method includes all possible quantum terms in the entropy current and 
if it agrees with the most fundamental quantum approach to dissipation, based on Zubarev 
non-equilibrium density operator \cite{Zubarev:1979}. Another very recent study
\cite{Bhadury:2020puc} studied the possible dissipative terms of the spin tensor in 
the relaxation time approximation.

\subsection{Hydrodynamic calculations}
\label{sec:hydrocalc}

The main goal of  hydrodynamic calculations is to provide the key input to the formula~\ref{basic}, that is the thermal vorticity at the freeze-out hypersurface. In principle,
the thermal vorticity field depends on the assumed initial conditions of the hydrodynamic
calculations, on the equation of state, on the hydrodynamic constitutive equations and 
on the freeze-out conditions. Nevertheless, different hydrodynamic calculations have 
provided similar results, which is reassuring regarding the robustness of theoretical computations
of polarization.

It is important to stress that polarization studies demand a 3+1D hydrodynamic simulation.
This is a crucial requirement because the components of the thermal vorticity driving the 
projection of the mean spin vector along the total angular momentum involve the gradients
of the longitudinal flow velocity, which are neglected by 2+1D codes. 

A common feature of all calculations is the fact that the values of thermal vorticity are,
on the average, sufficiently less than 1 so as to justify a linear approximation in the relation
between mean spin vector and thermal vorticity (see e.g. eq.~\ref{meanspin1}); this is
shown in the histogram in figure~\ref{vorthist}. Nevertheless, a role of quadratic corrections
cannot be excluded and it is yet to be studied.

The codes that have been used so far to calculate polarization based on formula~\ref{basic}
are few:
\begin{enumerate}
 \item {} A 3+1D {\em Particle in Cell} simulation of ideal relativistic hydrodynamics 
    \cite{Csernai:2013bqa}. All published calculations of polarization assume peculiar initial 
    conditions for heavy ion collisions, implying a non-vanishing initial vorticity.
 \item{} A 3+1D code implementing relativistic dissipative hydrodynamics, ECHO-QGP 
  \cite{DelZanna:2013eua} with initial conditions adjusted to reproduce the directed
   flow as a function of rapidity \cite{Bozek:2010bi}.
 \item{} A 3+1D code implementing relativistic dissipative hydrodynamics, vHLLE \cite{Karpenko:2013wva}  
  with initial state determined by means of a pre-stage of nucleonic collisions, and 
  including a post-hadronization rescattering stage, all adjusted to reproduce the basic 
  hadronic observables in relativistic heavy ion collisions, that is (pseudo)rapidity and 
  transverse momentum distributions and elliptic flow coefficients.
 \item{} A 3+1D code implementing relativistic dissipative hydrodynamics, CLVisc 
  \cite{Pang:2018zzo} with initial conditions provided by another transport-based simulation 
  package AMPT \cite{Lin:2004en}.
\end{enumerate}
Furthermore, many calculations of polarization in literature are based on the coarse-graining
of the output provided by the transport-based simulation code AMPT \cite{Lin:2004en} to 
obtain the thermal vorticity field in eq. \ref{basic}; we will refer to these calculations
as transport-hybrid.

Overall, while the global polarization is in excellent quantitative agreement with the hydrodynamic
calculations based on~\ref{basic}, the azimuthal dependence of the polarization along
angular momentum and the sign of the longitudinal component disagree with the data (c.f. sections~\ref{sec:SystematicsAt200GeV} and~\ref{sec:Longitudinal}).
These issues will be discussed later in Section~\ref{issues}.

\subsection{Effects of decays and rescattering}
\label{decays}

Most of the calculations presented in literature involve the {\em primary} $\Lambda$, i.e. 
those which are emitted from the freeze-out hypersurface. However, they are just a fraction 
of the measured $\Lambda$'s, about 25\% at \ssn = 200 GeV according to  statistical hadronization
model estimates~\cite{Wheaton:2004qb},
while most of them are  decay products of higher lying states, such 
as $\Sigma^0$, $\Sigma^*$, $\Xi$ etc. Those states are expected to be polarized as well,
according to the formula~\ref{basic} with the suitable spin-dependent coefficient (see e.g. 
eq.~\ref{meanspin2}), hence with the same momentum pattern as for the primary $\Lambda$'s.
The secondary $\Lambda$ from decays of polarized particles turns out to be polarized in turn 
and its polarization vector depends on the properties of the interaction responsible for the 
decay (strong, electromagnetic, weak) and on the polarization of the decaying particle. 
The formula for the global polarization inherited by the $\Lambda$'s
in several decay channels was obtained in ref.~\cite{Becattini:2016gvu} and its effect studied in
\cite{Karpenko:2016jyx}. While single channels involve a sizeable correction to the primary
polarization, the overall effect is small, of the order of 10\% or so. 
This result was confirmed by more detailed studies where the polarization transfer in 2-body decays
producing a $\Lambda$ hyperon was determined as a function of momentum \cite{Xia:2019fjf,Becattini:2019ntv}. Surprisingly, the combination of relative production rates of 
different hyperons, their decay branching ratios and the coefficients of the polarization transfer 
produce an accidental cancellation of the contribution of secondary $\Lambda$'s polarization
so that the dependence of polarization as a function of momentum is almost the same as predicted
for primary $\Lambda$'s alone \cite{Xia:2019fjf,Becattini:2019ntv}. 

While the contribution of secondary decays is under control, little is known about 
the effect of post-hadronization secondary hadronic scattering after the hydrodynamic 
motion ceases. In general, one would naturally expect an overall dilution of the primary 
polarization. However, it has been speculated~\cite{Voloshin:2004ha} that final-state 
hadronic rescattering could generate some polarization and a model was put forward in 
ref.~\cite{Barros:2005cy} showing that initially unpolarized hyperons in pA collisions 
can become polarized because of secondary interactions. However, the same model applied 
to AA yields a secondary polarization consistent with zero \cite{Barros:2007pt}. 

\subsection{Kinetic models}
\label{kinetic}

If, for some reason, spin degrees of freedom relax more slowly than momentum, local thermodynamic
equilibrium is not possibly a good approximation and the calculation of polarization becomes
more complicated. A possible substitute theoretical approach is kinetic theory. However, as 
has been mentioned, near the pseudo-critical temperature, the QGP is a strongly interacting 
system for which a kinetic description is dubious because the thermal wavelength of partons 
is comparable to their mean free path; particles interact so strongly that they are not 
free for most of their time. Notwithstanding, one may hope that kinetic theory provides a good 
approximation for the spin degrees of freedom if the spin-orbit coupling is weak. 
Recent estimates of the spin-flip rate in perturbative QCD imply, though, indicate a too large 
equilibration time \cite{Kapusta:2019sad} so that non-perturbative effects appear to be 
essential.

A formulation of relativistic kinetic theory with spin dates back to De Groot and collaborators
\cite{DeGroot:1980dk}, and it has been the subject of intense studies over the past few years. While
the development of a relativistic kinetic theory of massless fermions was motivated by the search
of the Chiral Magnetic Effect~\cite{Fukushima:2008xe,Li:2020dwr}, the corresponding theory for massive fermions 
is mostly motivated by the observation of polarization.
The goal of the relativistic kinetic theory of fermions is the study of the evolution of the
covariant Wigner function, which extends the notion of the phase space distribution function 
of relativistic Boltzmann equation. For free particles this reads:
\be\label{wigner}
   \widehat W(x,k)_{AB} = - \frac{1}{(2\pi)^4} \int \di^4 y \; \e^{-\ii k \cdot y}
    : \Psi_A (x-y/2) \Psibar_B (x+y/2) : 
\ee
where $\Psi$ is the Dirac field, $A,B$ are spinorial indices and $:$ denotes normal ordering; 
this definition should be changed to make it gauge invariant in quantum electrodynamics. 
Most recent studies aimed at a formulation of the covariant Wigner function kinetic equations 
in a background electromagnetic field \cite{Wang:2019moi,Hattori:2019ahi,Gao:2019znl,Weickgenannt:2019dks,Yang:2020hri} at some order in $\hbar$. A different approach 
was taken in ref.~\cite{Zhang:2019xya}, where the polarization rate was obtained including 
the spin degrees of freedom in the collisional rate of the relativistic Boltzmann equation. 

Kinetic theory with spin is in a theoretical development stage and has not yet produced 
stable numerical estimates of polarization in heavy ion collisions. However, important steps 
toward this goal have been recently made. In ref.~\cite{Li:2019qkf} an estimate of the 
evolution equation of the spin density matrix in perturbative QCD has been obtained. 
Computing tools are also being developed for the numerical solution of relativistic
kinetic equations \cite{Zhang:2019uor}. 

A sensitive issue of this approach is how to transfer the calculated polarization of partons 
to the hadrons, which is not relevant for the hydrodynamic-statistical model, see the discussion
at the beginning of this Section. More generally, there is a gap between the perturbative,
collisional quark-gluon stage and the hadronic final state which is highly non-trivial and
needs to be bridged.

\subsection{Spin tensor and spin potential}
\label{spintensor}

A very interesting theoretical issue concerned with the description of spin effects in relativistic 
fluids, is the possible physical separation between orbital and spin angular momentum. A similar 
discussion has been going on for several years in hadronic physics in connection with the proton 
spin studies \cite{Leader:2013jra}. A comprehensive introduction and discussion of the subject is 
beyond the scope of this work, we refer the reader to the specialized literature.

In Quantum Field Theory, the angular momentum current has in general two contributions: 
a so-called {\em orbital} part involving the stress-energy tensor and a {\em spin} part 
involving a rank three tensor $\spt^{\lambda,\mu\nu}$ called {\em spin tensor}. However, 
this separation seems to be unphysical and one can make a transformation of the stress-energy
and the spin tensor so as to make the current all orbital, obtaining the so-called Belinfante
stress-energy tensor, with the total angular momentum unchanged. This transformation is called 
pseudo-gauge transformation \cite{Hehl:1976vr} and it looks much like a gauge transformation in 
gauge field theories where the stress-energy and the spin tensor play the role of gauge 
potentials, while the total energy-momentum $P^\mu$ and angular momentum-boost $J^{\mu\nu}$
are gauge-invariant.
The question is whether an observation of a polarization in the QGP breaks pseudo-gauge 
invariance, making it possible to single out a specific spin tensor. This would be obviously
a breakthrough with remarkable consequences, as it would have an impact on fundamental physics,
such as relativistic gravity theories. 

Indeed, the first derivation of the formula~\ref{basic} made use of a specific spin tensor and
this has led to some confusion, even in the original paper \cite{Becattini:2013fla}. In fact,
it was later observed \cite{Florkowski:2017dyn} that the resulting expression of the polarization
is the same regardless of the spin tensor used, among the most common choices. It has recently
become clear that the definition of spin density matrix and of the spin vector 
\cite{Florkowski:2018fap,Florkowski:2019gio} in Quantum Field Theory do not indeed require 
any angular momentum or spin operator, just the density operator and creation-destruction operators \cite{Becattini:2019ntv}; so, their expressions are completely independent of the spin 
tensor. In fact, the {\em mean value} of the polarization may depend on the spin tensor, insofar as
the density operator does. At global thermodynamic equilibrium, the density operator~\ref{densop2} 
is manifestly independent of the spin tensor because only the total angular momentum appears, 
but in the case of {\em local} thermodynamic equilibrium, the density operator~\ref{ledo} is 
not invariant under a pseudo-gauge transformation \cite{Becattini:2018duy}. Then, in principle, 
one might be able to distinguish between two spin tensors by measuring the polarization. Of course, 
this is a principle statement because, in practice, there are many uncertainties limiting the 
accuracy of the theoretical predictions (e.g. the hydrodynamic initial conditions) and it is not 
clear yet to what extent the measurements could solve the issue. 

The inclusion of the spin tensor in relativistic hydrodynamics has been explored in some detail 
by W. Florkowski {\it et al.} in a series of papers \cite{Florkowski:2017ruc},\cite{Florkowski:2018fap} 
and a first hydrodynamic calculation of polarization presented in a simplified boost-invariant scenario \cite{Florkowski:2019qdp}. 
As far as the heavy ion phenomenology is concerned, a general comment is in order for the spin 
tensor scenario: an extended version of relativistic hydrodynamics requires six additional  
fields (the anti-symmetric spin potential $\Omega_{\mu\nu}$) which in turn need six additional 
initial and boundary conditions, which are completely unknown in nuclear collisions. 
Polarization measurements could then be used to adjust them, but 
this would strongly reduce the probing power of polarization in all other regards.

\subsection{Contribution of the electro-magnetic field}
\label{sec:em}

As has been mentioned, an important feature of the statistical-thermodynamic approach is that
polarization is independent of the charge of the particles for a charge-neutral fluid. 
This has been confirmed by the measurements, which essentially find the same 
magnitude and sign for $\Lambda$ and $\bar\Lambda$ polarization (see figure~\ref{fig:RootsDependence}
later in this work). Indeed, for a fluid with some charge current, a difference in the polarization 
of particle and anti-particle is encoded in the Fermi-Dirac distributions in eq.~\ref{basic} 
in that the e.g. baryon chemical potential 
is larger at lower energy, favouring the $\bar\Lambda$'s polarization through the factor $n_F(1-n_F)$ 
in the numerator~\cite{Fang:2016vpj}. However, the known values of baryon chemical potential/temperature 
ratios at the relevant collision energies imply a much smaller difference in the polarization than
observed.

A possible source of particle-antiparticle polarization splitting is the electro-magnetic field, 
which would lead - at local equilibrium - to a modification of the formula~\ref{basic} with
thermal vorticity $\varpi^{\mu\nu}$ replaced by \cite{Becattini:2016gvu}:
\be\label{emfield}
    \varpi_{\rho\sigma} \to \varpi_{\rho\sigma} + \frac{\mu}{S} F_{\rho\sigma}
\ee
with $\mu$ the particle magnetic moment. Indeed, in peripheral heavy ion collisions 
  a large electro-magnetic field is present at the collision time which 
may steer the spin vector of $\Lambda$ and $\bar\Lambda$ and lead to a splitting of polarization, 
their magnetic moments being opposite.  

Therefore, the polarization splitting might be taken advantage of to determine the magnitude of the 
electro-magnetic field at the freeze-out (or earlier if the relaxation time is not small)~\cite{Becattini:2016gvu} 
or its lifetime~\cite{Guo:2019joy}. Pinning down the electro-magnetic field would be a very important 
achievement in the search of local parity violation in relativistic heavy-ion collisions \cite{Kharzeev:1998kz} 
through the so-called Chiral Magnetic Effect \cite{Fukushima:2008xe,Li:2020dwr}. However, 
alternative explanations of the splitting have been proposed and this feature needs to be explored experimentally and theoretically. We will return to this in section~\ref{sec:OpenQuestionSplitting}.

\section{POLARIZATION IN RELATIVISTIC HEAVY ION COLLISIONS: OBSERVATIONS} 
\label{sec:Observations}

As of this writing, there is only a handful of measurements of spin polarization
  in relativistic heavy ion collisions.
These measurements require excellent tracking and vertex resolution in the region of interest
  (typically midrapidity); large coverage and good particle identification to measure decay products;
  high statistics to measure relatively small correlation signals; and a suite of detectors
  to correlate forward-rapidity momentum anisotropies with midrapidity decay topologies.
Several such experiments exist today, and more will soon be commissioned.
The initial measurements described here will eventually be part of a fuller set of mapped systematics.

\subsection{Measuring polarization}
\label{sec:MeasuringPolarization}

If spin is locally equilibrated, as we have discussed, all
  hadrons with spin will be polarized.
However, while polarimeters~\cite{Jinnouchi:2003cp} 
  may directly detect the polarization of particles in very clean environments, their use
  is infeasible in a final state involving thousands of hadrons.

Recording the debris from the midrapidity region in a heavy ion collision usually involves large tracking
  systems~\citep[e.g.][]{Anderson:2003ur}.
A particle's polarization may be determined by the topology of its decay into charged particles,
  if the angular distribution of daughters' momenta is related to the spin direction of the parent.
  
For weak parity-violating hyperon decays with spin and parity 
  $\tfrac{1}{2}^{+}\rightarrow\tfrac{1}{2}^{+}+0^{-}$,
  the daughter baryon is emitted preferentially in the
  direction of the polarization vector ($\mathbf{P}^{*}_{H}$) of the parent,
  as~\cite{PhysRev.108.1645}
\begin{equation}
\label{eq:HyperonDecay}
\frac{{\rm d}N}{{\rm d}\Omega^*}
=\frac{1}{4\pi}\left(1+\alpha_{H}\mathbf{P}^{*}_{H}\cdot\mathbf{\hat{p}}^{*}_{D}\right) 
=\frac{1}{4\pi}\left(1+\alpha_{H}\cos\xi^*\right) 
,
\end{equation}
where $\mathbf{\hat{p}}^{*}_{D}$ is a unit vector pointing in the direction of the daughter baryon momentum, and
$\xi^*$ is the angle between the $\mathbf{\hat{p}}^*_D$ and the polarization direction.
Here and throughout, an asterisk ($^*$) denotes a quantity as measured in the
  rest frame of the decaying parent.
The decay parameter $\alpha_{H}$ depends on the hyperon species~\cite{PhysRevD.98.030001}.
  
The general task to extract polarization from experimental data is to identify
  a potential direction, say $\mathbf{\hat{n}}$ (specific examples discussed below).
The ensemble-averaged projection of the daughter baryon's momentum along $\mathbf{\hat{n}}$ 
  gives the projection of $\mathbf{P}$:
  \begin{equation}
   \label{eq:HyperonDecay2}
      \left\langle\mathbf{\hat{p}}_{D}^*\cdot\mathbf{\hat{n}}\right\rangle = \frac{\alpha_H}{3}\mathbf{P}^*_H\cdot\mathbf{\hat{n}} .
  \end{equation}

First measurements~\cite{Abelev:2007zk,STAR:2017ckg,Adam:2018ivw,Adam:2019srw,Acharya:2019ryw}
  of polarization in relativistic heavy ion collisions have
  used $\Lambda\rightarrow p+\pi^-$ ($\overline{\Lambda}\rightarrow\overline{p}+\pi^+$) decays.
The decay parameter for an antiparticle is expected and 
  observed~\cite{PhysRevD.98.030001,Ablikim:2018zay} to be of equal magnitude and opposite sign
  of the corresponding particle within measurement uncertainties.\footnote{Until very recently, the accepted world average value has been~\cite{PhysRevD.98.030001} 
  $\alpha_{\Lambda}=0.642\pm0.013$.
However, a recent measurement~\cite{Ablikim:2018zay} by the BES-III Collaboration reports
    $\alpha_{\Lambda}=0.750\pm0.009\pm0.004$, a discrepancy of about 10$\sigma$.
Although the source of this large discrepancy not entirely clear, in its online 2019 update, the Particle Data Group adopted this new value.  Therefore, we have decided to scale all reported polarizations to reflect the BESIII value.}
  
Polarization of other hadronic species may also also be measured, in principle.
The reduced efficiency associated with identifying two displaced vertices, as well as the reduced yield of doubly strange baryons
  makes using $\Xi^-$ ($\alpha_{\Xi^{-}\rightarrow\Lambda+\pi^{-}}=-0.458$) more difficult.
The neutral decay of $\Xi^0$ ($\alpha_{\Xi^{0}\rightarrow\Lambda+\pi^{0}}=-0.406$) is more difficult still.
Relatively low production rates~\cite{BraunMunzinger:2001ip} and very small $\alpha_{\Omega}$ values~\cite{PhysRevD.98.030001}
  strongly disfavor the use of triply-strange $\Omega$ baryons.

For spin-1/2 particles, polarization is entirely described by 
the mean spin vector, which has been extensively discussed in this work. For particles with
spin $>1/2$, a full description of the polarization state requires more quantities; in practice,
one should quote the full spin density matrix $\Theta_{rs}(p)$ (see Section \ref{theory}).
Particularly, for spin 1 particles, a quantity independent of the mean spin vector related
to the polarization state is the so-called {\em alignment} \cite{Leader:2001gr}:
$$
A = \Theta_{00}(p)-1/3
$$
A randomly-oriented ensemble would have $\Theta_{00}=\tfrac{1}{3}$, hence vanishing $A$;
 a value $\Theta_{00} \ne \tfrac{1}{3}$ indicates spin alignment, though by symmetry
  it is impossible to distinguish the sign in $\left\langle\vec{S}\right\rangle\parallel\hat{n}$.
The 2-particle decay topology of a vector meson is related to the alignment according 
to~\cite{Schilling:1969um}:
\begin{equation}
    \label{eq:SpinAlignment}
    \frac{{\rm d}N}{{\rm d}\cos\xi^*} = \frac{3}{4}\left[1-\Theta_{00}+\left(3\Theta_{00}-1\right)\cos^2\xi^*\right] ,
\end{equation}
where $\xi^*$ is defined as in equation~\ref{eq:HyperonDecay}. At local thermodynamic
equilibriu, $\tfrac{1}{3}-\rho_{00}$ is quadratic in thermal vorticity to first order, as 
mentioned in section~\ref{theory}.
  
Thus far, the first measurements of global spin alignment of vector mesons 
in heavy ion collisions are difficult to understand in a consistent picture. We discuss these 
in subsection \ref{sec:align}, and focus here on hyperon polarization.

\subsection{Global hyperon polarization - observation}
\label{sec:GlobalData}

By symmetry, the average vorticity of the QGP fireball must point in the direction
  of the fireball's angular momentum $\vec{J}_{\rm QGP}$, and on average
  $\vec{J}_{\rm QGP}\parallel\vec{J}_{\rm sys}$ (c.f. figure~\ref{fig:CollisionSketch}).
Similarly, even without appealing to a connection to vorticity, when averaged
  over all particles, symmetry demands an average (over all emitted particles) polarization aligned with
  $\hat{J}_{\rm sys}$.
In the current context, the "global polarization" of a subset of particles 
  refers to the use of $\hat{n}=\hat{J}_{\rm sys}$ in equation~\ref{eq:HyperonDecay2}.

\begin{figure}[t]
\begin{minipage}{0.35\textwidth}
  \includegraphics[width=\textwidth]{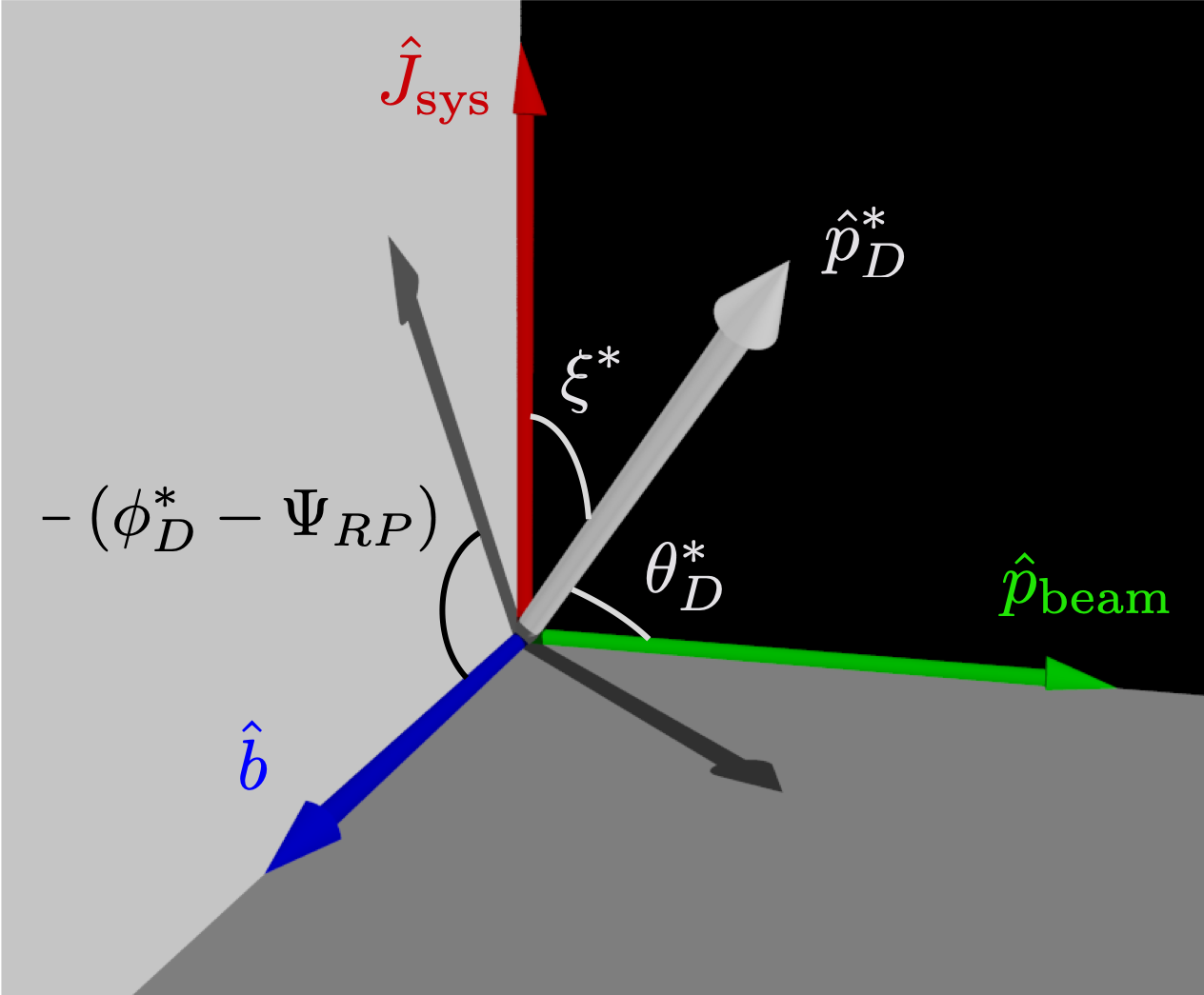}
\end{minipage}
\begin{minipage}{0.65\textwidth}
  \includegraphics[width=\textwidth]{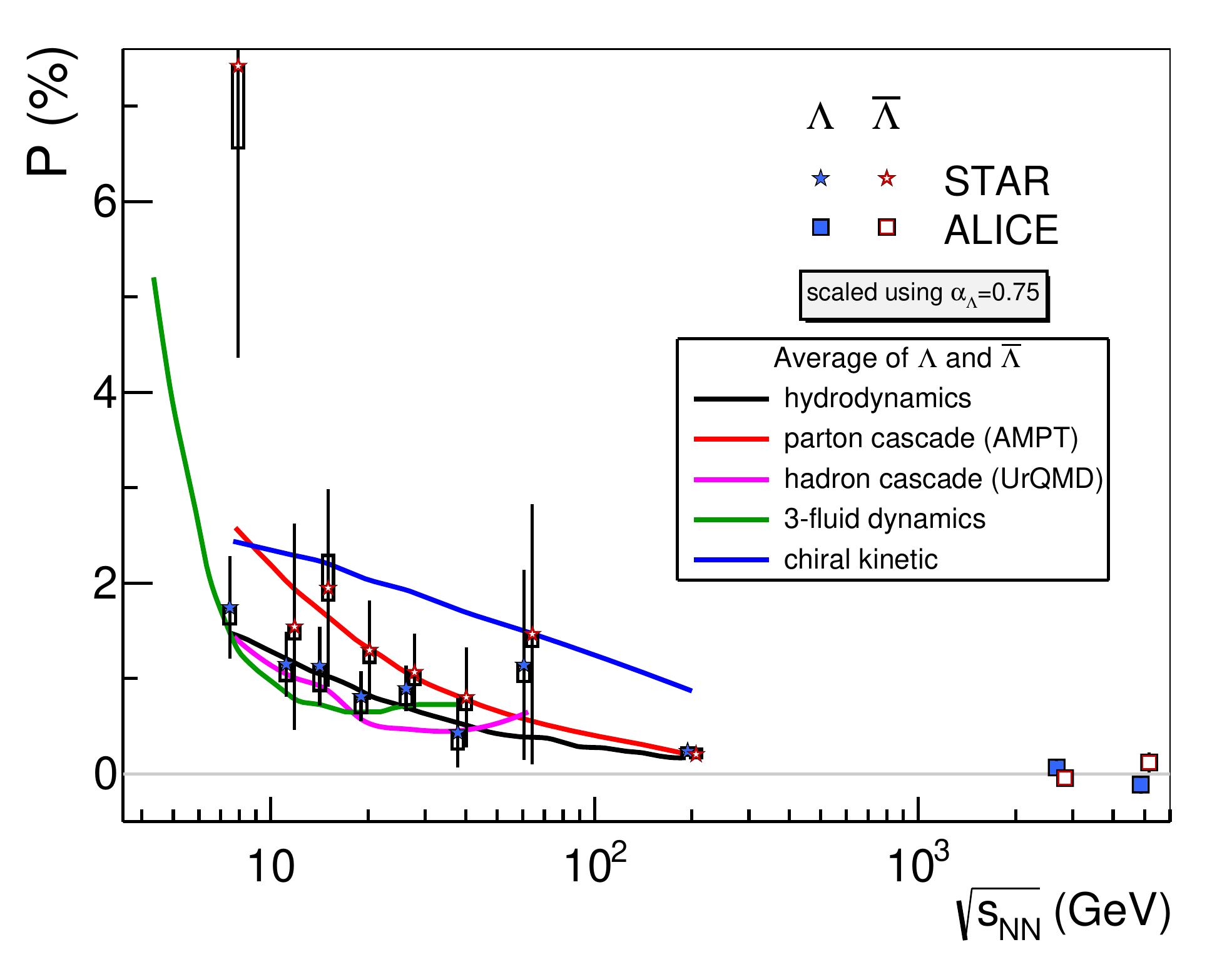}
\end{minipage}
\caption{Left: \label{fig:VariousAngles}
The vectors and angles involved in an analysis of hyperon polarization along the angular momentum of the
  collision are shown.
In the lab coordinate system (not shown), the azimuthal angle of $\hat{b}$ is defined to be $\Psi_{RP}$.
Thus, the angle between the projection of $\hat{p}^*_D$ and $\hat{b}$ is
  $\phi^*_D-\Psi_{RP}$.
The minus sign on the angle indicated arises from the fact that azimuthal angles
  are measured counterclockwise about the beam axis.
  \newline
  Right: \label{fig:RootsDependence} The energy dependence of $\Lambda$ and $\overline{\Lambda}$ global
polarization at mid-rapidity from mid-central Au+Au (20-50\%) or Pb+Pb (15-50\%) collisions.
Data~\cite{STAR:2017ckg,Abelev:2007zk,Adam:2018ivw,Acharya:2019ryw} are compared to
polarization simulations of viscous hydrodynamics~\cite{Karpenko:2016jyx};
partonic transport~\cite{Li:2017slc}; hadronic transport~\cite{Vitiuk:2019rfv};
chiral-kinetic transport plus coalescence~\cite{Sun:2017xhx};
and a three-fluid hydro model applicable at lower energies~\cite{Ivanov:2019ern}.
Experimental data points have been corrected for the recent change in $\alpha_\Lambda$, as discussed
  in section~\ref{sec:MeasuringPolarization}.
For~\cite{Karpenko:2016jyx} and~\cite{Li:2017slc}, the values shown represent both
  primary and feed-down hyperons~\citep[c.f][]{Becattini:2016gvu}.
See text for details.}
\end{figure}


As discussed in section~\ref{sec:Conventions}, the momentum-space anisotropy of particle emission is
  used~\cite{Poskanzer:1998yz} to extract
  an event plane angle $\Psi_{EP,1}$ which approximates the reaction plane with some finite resolution.
Standard methods have been developed~\cite{Poskanzer:1998yz} to correct 
  for the effects of this resolution on measured asymmetries in the emission pattern about
  the beam axis, so it is convenient to rewrite equation~\ref{eq:HyperonDecay2} 
  as\footnote{See reference~\cite{LisaSpringer:2020} for a discussion of significant experimental
  challenges to perform the average in equation~\ref{eq:Phat1}}\cite{Abelev:2007zk}
  \begin{equation}
    \label{eq:Phat1}
      P_{H,\hat{J}}
      =\frac{3}{\alpha_H}\left\langle\mathbf{\hat{p}^*_D}\cdot\left(\mathbf{\hat{b}}\times\mathbf{\hat{p}_{\rm beam}}\right)\right\rangle
      = \frac{3}{\alpha_H}\left\langle\cos\xi^*\right\rangle
      = -\frac{3}{\alpha_H}\left\langle\sin\left(\phi^*_D-\Psi_{\rm RP}\right)\sin\theta^*_D\right\rangle .
  \end{equation}
Here, $\phi^*_D$ and $\theta^*_D$ are the angles between the daughter momentum and $\mathbf{\hat{b}}$ and
  $\mathbf{\hat{p}}_{\rm beam}$, respectively,
  and in the last step, a trigonometric relationship between the angles is used.
These angles are shown in figure~\ref{fig:VariousAngles}.

Integrating\footnote{\label{footnote:ThetaIntegration}Detectors
  in which $\Lambda$s are reconstructed usually do not measure the charged daughters
  at very forward angles at collider energies.
  Corrections~\cite{Abelev:2007zk,Lan:2017nye} on order~$\sim3\%$~\cite{Upsal:2018phd}
  are applied in
  order to account for this.} over polar angle $\theta^*$
\begin{equation}
    \label{eq:ExperimentalQuantity}
    P_{H,\hat{J}}
       =-\frac{8}{\pi\alpha_H}\left\langle\sin\left(\phi^*_D-\Psi_{\rm RP}\right)\right\rangle
              =-\frac{8}{\pi\alpha_H R_{\rm EP}^{(1)}}\left\langle\sin\left(\phi^*_D-\Psi_{\rm EP,1}\right)\right\rangle ,
\end{equation}
where in the final step, the experimentally-determined event plane angle replaces the
   reaction plane angle, accounting for the resolution with a calculable correction factor 
   $R_{\rm EP}^{(1)}$~\cite{Poskanzer:1998yz}.

The resolution with which $\hat{J}_{\rm sys}$ is measured is critical.
Polarization affects daughter anisotropies only at the few percent level, and
  statistical uncertainties can dominate experimental results.
Using equation~\ref{eq:ExperimentalQuantity}. the statistical uncertainty on polarization goes as
  \begin{equation}
      \delta P_{H,\hat{J}} \sim \left(R_{\rm EP}^{(1)}\cdot\sqrt{N_{H}}\right)^{-1} ,
  \end{equation}
where $N_{H}$ is the total number of hyperons analyzed in the dataset.
This dependence is generically true for any measurement that involves correlation with the
  first order event plane or $\hat{J}_{\rm sys}$.
Increasing the resolution by a factor of two~\cite{Adams:2019fpo} thus decreases the required duration of an
  experimental campaign four-fold.

Figure~\ref{fig:RootsDependence} shows the world dataset of $P_{\Lambda,\hat{J}}$
  and $P_{\overline{\Lambda},\hat{J}}$ as a function of collision energy for semi-peripheral
  collisions.
As discussed in section~\ref{sec:MeasuringPolarization}, the recent change in accepted value
  for $\alpha_{\Lambda}$ requires a rescaling of the published experimental values.
From a maximum of $\sim1.5\%$ at $\sqrt{s_{NN}}=7.7$~GeV, the polarizations fall
  smoothly\footnote{While eye-catching, the value of $P_{\overline{\Lambda},\hat{J}}=(7.4\pm3.1)\%$
   at the lowest energy is less than $2\sigma$ above the general systematics and is marginally significant.}
  with energy.
At LHC energies, they vanish within experimental uncertainties.

Strikingly, all available hydrodynamic and transport calculations reproduce the
  observations in sign, magnitude and energy dependence, as discussed in section~\ref{sec:hydrocalc}. 
This is nontrivial; since they all use formula~\ref{basic} it means that they all 
  predict a very similar thermal vorticity field.
These models have been to some extent "tuned" 
  to reproduce earlier observations such as anisotropic flow~\cite{Heinz:2013th}, which is sensitive to the
  bulk motion {\it of} fluid cells from which particles emerge; it
  is thus satisfying that they produce similar and correct predictions for this more sensitive observable.

Clearly, $|\vec{J}_{\rm sys}|$ increases with increasing $\sqrt{s_{NN}}$,
  and transport calculations~\cite{Jiang:2016woz,Baznat:2015eca}
  predict that about 20\% of this
  angular momentum is transferred to the QGP fireball.
While some early calculations~\cite{Gao:2007bc} predicted an
  increased polarization at high collision energies, a strongly
  decreasing trend is produced by most
  hydrodynamic~\cite{Karpenko:2016jyx,Xie:2017upb}
  and transport-hybrid codes in which the thermal vorticity field
  is obtained through a coarse-graining 
  procedure~\cite{Jiang:2016woz,Deng:2016gyh,Xia:2018tes,Wei:2018zfb,Wu:2019eyi}.
  
Driving mechanisms may include 
  increased temperature~\cite{Csernai:2014ywa} at increased $\sqrt{s_{NN}}$; increases in evolution
  timescale~\cite{Karpenko:2016jyx,Sun:2017xhx}; vorticity migrating to forward 
  rapidity~\cite{Betz:2007kg,Jiang:2016woz,Ivanov:2019ern}, perhaps due to reduced baryon stopping / increased transparency at high energy~\cite{Karpenko:2016jyx}; an increased
  fluid moment of inertia due to increased mass-energy~\cite{Jiang:2016woz}; reduced longitudinal fluctuations and boost-invariance at high energy~\cite{Pang:2016igs}.

In addition to the overall energy dependence, the data in figure~\ref{fig:RootsDependence}
  suggests a fine splitting between particles and antiparticles at low $\sqrt{s_{NN}}$.
While statistically not significant at
  any given energy, very important physical effects are predicted to manifest in 
  $P_{\overline{\Lambda},\hat{J}}>P_{\Lambda,\hat{J}}$, as we discuss
  in section~\ref{sec:OpenQuestionSplitting}.

Even as we note possible differences between the polarizations of $\Lambda$ and $\overline{\Lambda}$, it
  is clear that to good approximation they are the same, even at the lowest energies, suggesting similar average
  vorticity of the cells from which they arise.
This is remarkable, in light of the fact that the directed flow of
  these particles diverge strongly~\cite{Adamczyk:2017nxg}
  as the energy is reduced below $\sqrt{s_{NN}}\approx20$~GeV, even taking 
  opposite signs at midrapidity.
In the hydrodynamic paradigm, directed flow~\cite{Poskanzer:1998yz}, essentially the sidewards push of forward-going particles (c.f.  figure~\ref{fig:CollisionSketch})
  reflects the anisotropy of the bulk fluid velocity about the $\hat{J}$ axis at a large scale.
Meanwhile, global polarization reflects rotational flow structure about
  $\hat{J}$ at a more local scale.
There may be a coupling in a hydrodynamic picture~\cite{Csernai:2011qq,Csernai:2014ywa,Becattini:2015ska}.
Whether there is a tension here is unclear, though a
  three-fluid hydrodynamic code is able to approximately reproduce proton and antiproton flow~\cite{Ivanov:2014ioa} and $\Lambda$ polarization~\cite{Ivanov:2019ern}.

\begin{figure}[t]
\centering
  \begin{minipage}[b]{0.495\textwidth}
    \includegraphics[width=\textwidth]{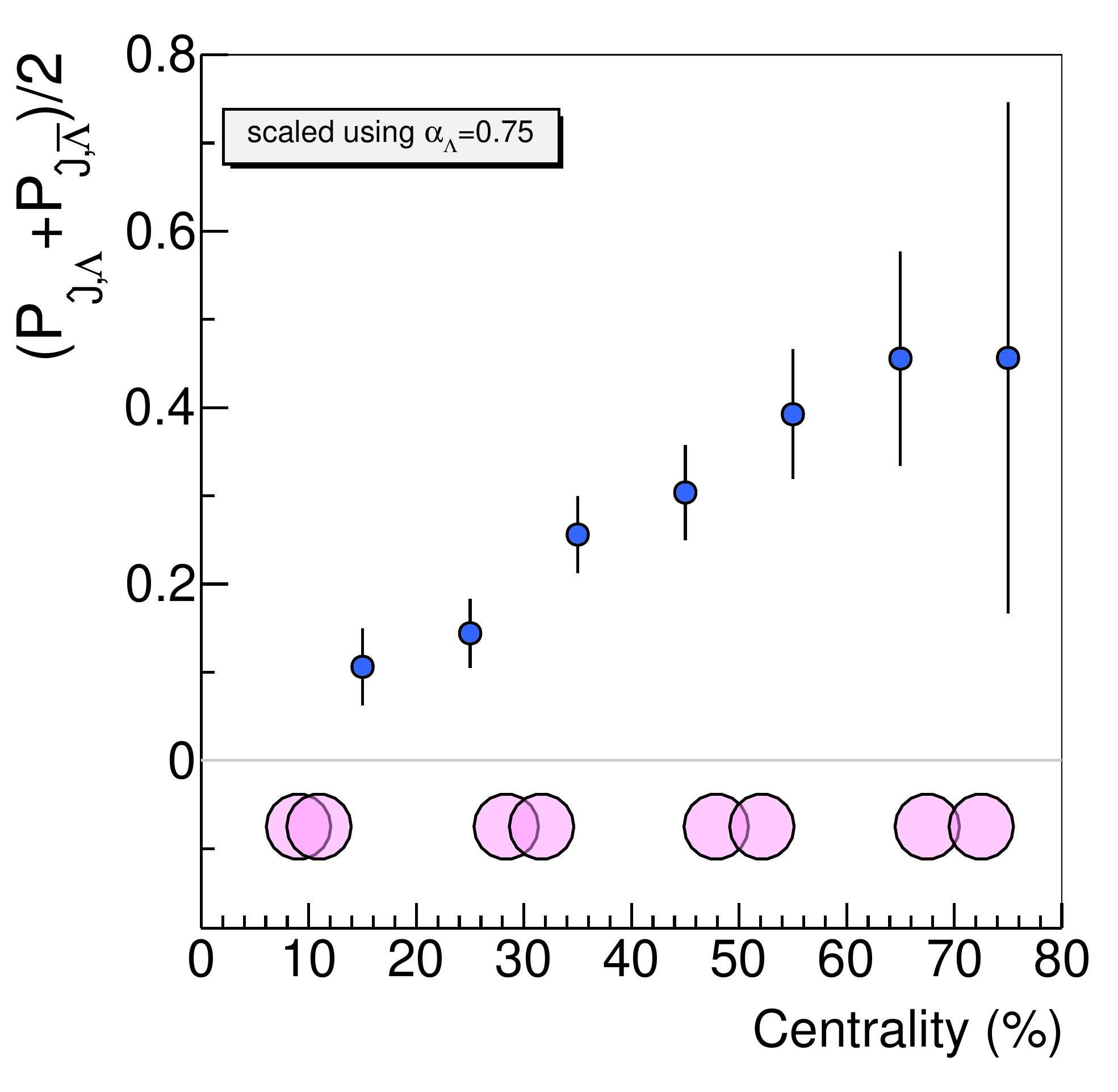}
  \end{minipage}
  \begin{minipage}[b]{0.495\textwidth}
    \includegraphics[width=\textwidth]{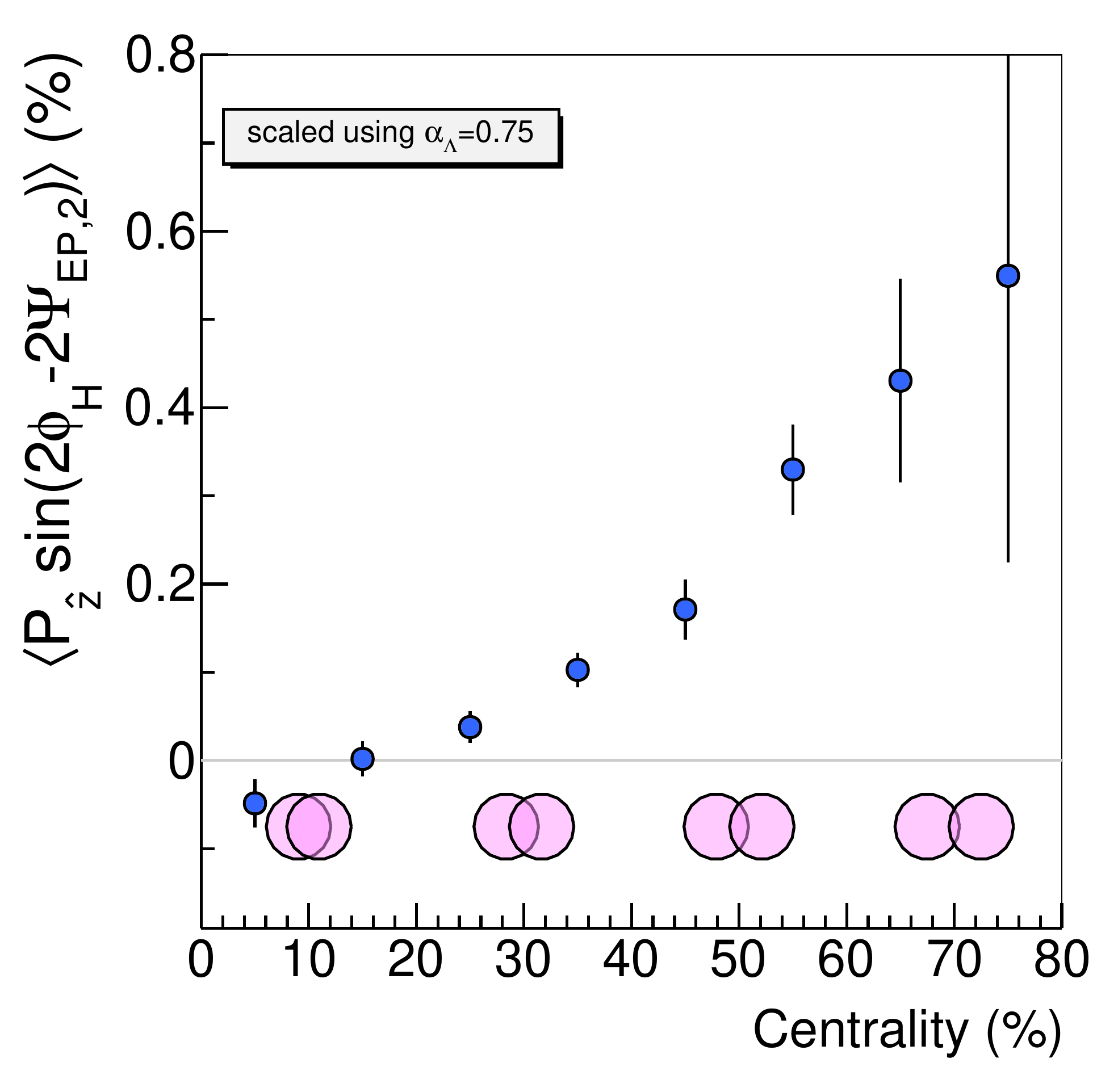}
  \end{minipage}
\caption{\label{fig:CentralityDependence}
The centrality dependence of hyperon (average of $\Lambda$ and $\overline{\Lambda}$)
  polarization in 200 GeV Au+Au collisions.
As in figure~\ref{fig:RootsDependence}, published data have been rescaled to reflect the new accepted
  value of $\alpha_\Lambda$.
Cartoons at the bottom of each panel roughly sketch the geometry of the overlap region for a given centrality.
Left panel: Global polarization~\cite{Adam:2018ivw}.
Right panel: Second-order oscillation amplitude of the longitudinal polarization~\cite{Adam:2019srw}.}
\end{figure}

\subsection{Global and local polarization at $\sqrt{s_{NN}}=200$~GeV}
\label{sec:SystematicsAt200GeV}
\label{sec:Longitudinal}

Systematic studies of the dependence of $P_{\hat{J},H}$ in Au+Au collisions have so far only been possible
  at $\sqrt{s_{\rm NN}}=200$~GeV~\cite{Adam:2018ivw}.
Statistics are poor at low energies, while at higher energies, the signal itself vanishes.
More detailed measurements can provide stringent
 challenges to theoretical models and may provide new insight.
In $\sqrt{s_{NN}}=200$~GeV collisions, polarizations of $\Lambda$ and $\overline{\Lambda}$ are
  identical within uncertainties, so here we discuss their average.

In figure~\ref{fig:RootsDependence}, global hyperon polarization was shown for collisions with
  centrality of 20-50\% (c.f. section~\ref{sec:Conventions}), corresponding to $|\vec{b}|\approx7-11$~fm.
Figure~\ref{fig:CentralityDependence} shows the centrality 
  dependence.
Both the global polarization and the oscillation of the longitudinal local polarization (c.f. section~\ref{sec:Longitudinal})
  increase monotonically with impact parameter, as expected for a phenomenon driven by bulk mechanical angular momentum; this is in agreement with transport-hybrid calculations~\cite{Jiang:2016woz}.
  
\begin{figure}[t]
\begin{minipage}{0.5\textwidth}
  \includegraphics[width=\textwidth]{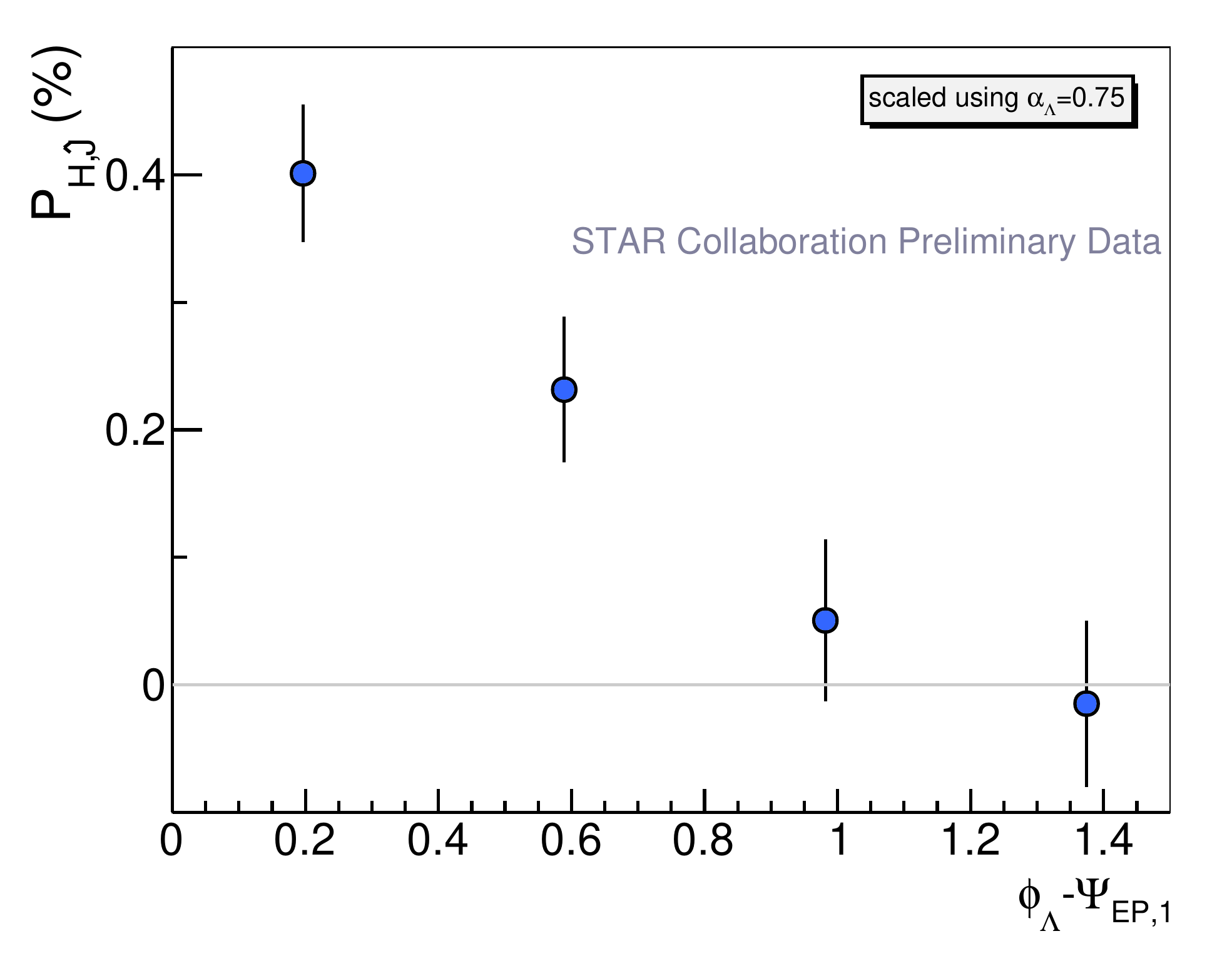}
\end{minipage}
\begin{minipage}{0.48\textwidth}
  \includegraphics[width=\textwidth]{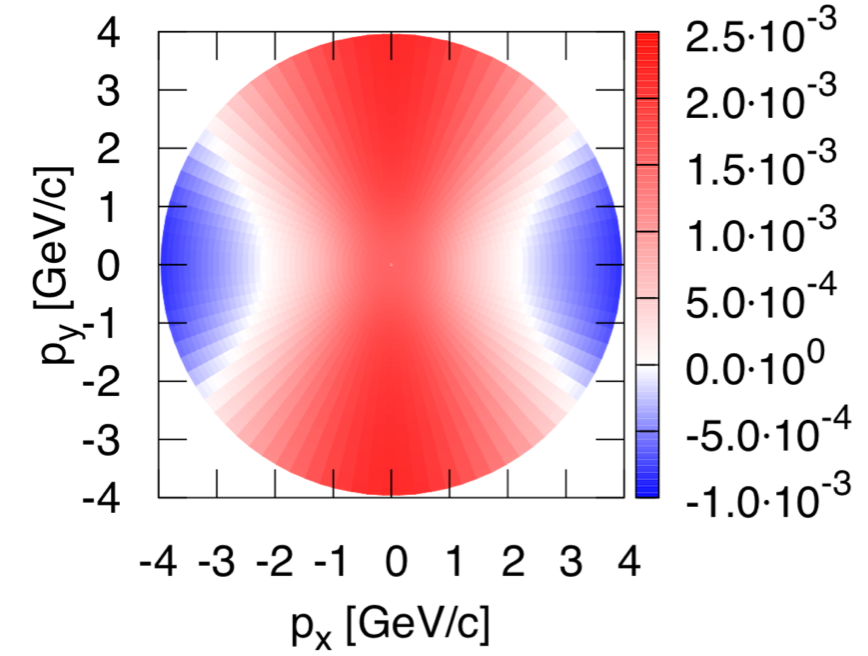}
\end{minipage}
\caption{\label{fig:GlobalVersusPhi}
Left: Preliminary results~\cite{Niida:2018hfw} from the STAR collaboration for the global polarization of
  $\Lambda$ and $\overline{\Lambda}$ as a function of hyperon emission angle relative to the event plane,
  for mid-central Au+Au collisions at $\sqrt{s_{NN}}=200$~GeV.
As in figure~\ref{fig:RootsDependence}, published data have been rescaled to reflect the new accepted
  value of $\alpha_\Lambda$.
Right: Hydrodynamic calculations~\cite{Becattini:2015ska} of
  $P_{\hat{J}}$ in the transverse momentum plane, for the same colliding system.
}
\end{figure}

The "global" polarization-- i.e. integrated over all particles at midrapidity-- is non-zero and aligned with an event-specific
  direction.\footnote{This is in strong contrast to the well-known phenomenon~\cite{Bunce:1976yb,Heller:1978ty} in p+p and p+A
  collisions, in which $\Lambda$ (but not $\overline{\Lambda}$, for unclear reason) hyperons emitted at very forward angles are polarized along their {\it production} plane, spanned by $\vec{p}_{\Lambda}\times\vec{p}_{\rm beam}$.  This effect
  is rapidity-odd, vanishing at midrapidity.
  In principle, convolution of the production-plane polarization with finite directed flow~\cite{Poskanzer:1998yz} could produce a global effect.  However, in practice, this effect is much smaller than those we discuss here~\cite{Becattini:2013vja}.}  
Momentum-differential ("local") polarization structures, in the local equilibrium picture,
are more sensitive to the thermal vorticity variations as a function of space and time, convoluted
  with flow-driven space-momentum correlations.
First measurements~\cite{Adam:2018ivw} report
  $P_{\hat{J},\Lambda/\overline{\Lambda}}$ to be independent
  of transverse momentum for $p_T\lesssim2$~GeV/c, in agreement with hydrodynamic predictions~\cite{Becattini:2015ska,Becattini:2017gcx} when realistic initial conditions are used.
It was also seen~\cite{Adam:2018ivw} to be independent of pseudo-rapidity, though only
  a limited range, $|\eta|<1$ could be explored.
As we discuss in section~\ref{sec:OpenQuestions}, several theories
  suggest there is much to be learned at forward rapidity.

A recurring theme in heavy ion physics has been that azimuthal dependencies often present surprises and the
  opportunity for new physical insight.
The same may well be true for polarization.
Figure~\ref{fig:GlobalVersusPhi} shows 
  preliminary data from the STAR collaboration~\cite{Niida:2018hfw} suggesting that $P_{\hat{J},\Lambda\&\overline{\Lambda}}$
  is significantly stronger for particles emitted perpendicular to $\hat{J}_{sys}$ ($|\phi_{\Lambda}-\Psi_{\rm RP}|=\pi/2$)
  than for $\hat{p}_{\Lambda}\parallel\hat{J}$.
Indeed, $P_{\hat{J}}$ may vanish for hyperons emitted out of the reaction plane.
This stands in contradiction to rather robust predictions
 of  hydrodynamic~\cite{Becattini:2013vja,Becattini:2015ska,Karpenko:2016jyx,Xie:2016fjj,Xie:2017upb}
and coarse-grained transport~\cite{Jiang:2016woz,Xia:2018tes,Wei:2018zfb,Wu:2019eyi} calculations,
 one of which is shown on the
  right panel of the figure, which predict precisely the opposite dependence.
If the STAR results are confirmed in a final analysis, this represents a nontrivial challenge to 
the theory.


\begin{figure}[t]
\begin{minipage}{0.5\textwidth}
  \includegraphics[width=\textwidth]{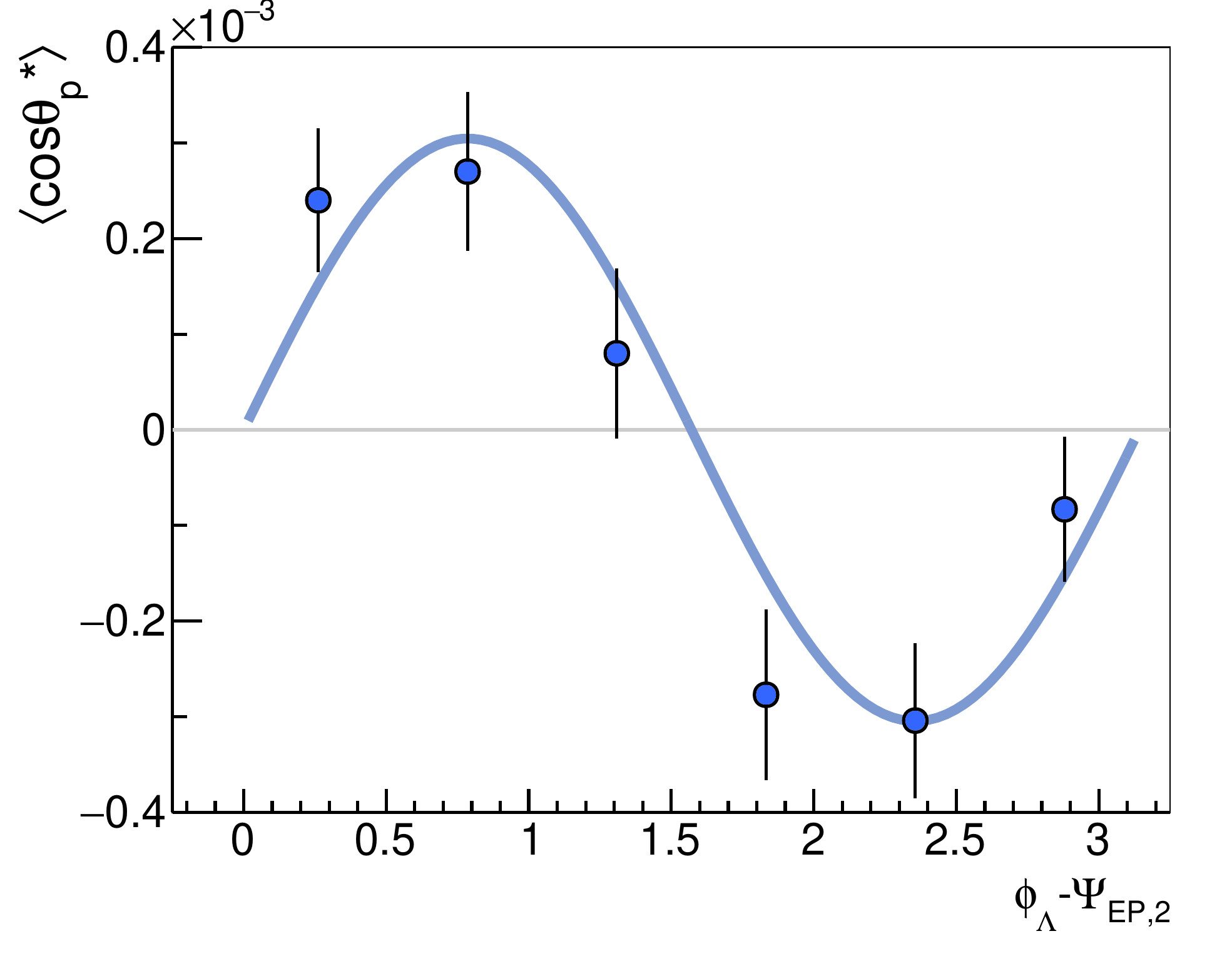}
\end{minipage}
\begin{minipage}{0.48\textwidth}
  \includegraphics[width=\textwidth]{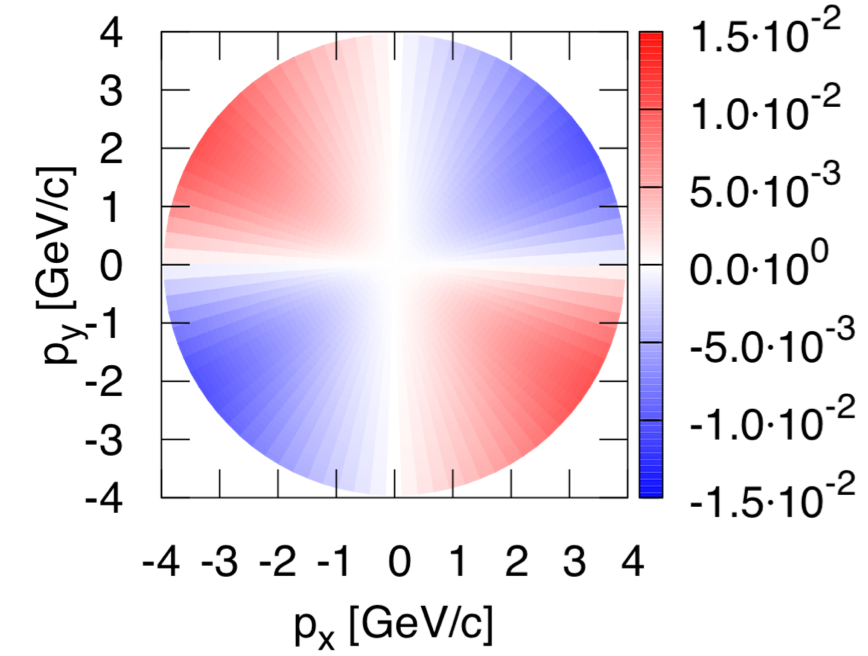}
\end{minipage}
\caption{\label{fig:LongitudinalVersusPhi}
$\left\langle\cos\theta^*_p\right\rangle$ for 20-60\% centrality Au+Au collisions at $\sqrt{s_{NN}}=200$~GeV,
  as a function of hyperon emission angle relative to the event plane~\cite{Adam:2019srw}.
Small detector effects (see footnote~\ref{footnote:ThetaIntegration}) and event-plane resolution effects have not been corrected for, in this figure.
A sinusoidal curve is drawn to guide the eye.
Right: Hydrodynamic calculations~\cite{Becattini:2015ska} of
  $P_{\hat{z}}$ in the transverse momentum plane, for the same colliding system.
}
\end{figure}

By symmetry, polarization components perpendicular to $\hat{J}_{\rm sys}$ must vanish,
 when averaging over all momenta. Locally in momentum space, however, these 
 components are allowed to be non vanishing. Particularly, there can be non-vanishing
 values oscillating as a function of the azimuthal emission angle $\phi_H$ over the 
 transverse plane with a typical quadrupolar pattern. Hydrodynamic~\cite{Becattini:2015ska} 
 and transport calculations~\cite{Xia:2018tes} 
 predict the sign and the magnitude of these oscillations. Here, $\hat{n}=\hat{p}_{\rm beam}$ 
 in equation~\ref{eq:HyperonDecay2} so $\xi^*_D=\theta^*_D$, the 
 polar angle of the daughter in the hyperon frame; c.f. figure~\ref{fig:VariousAngles}.
  
Hydrodynamic~\cite{Becattini:2015ska,Karpenko:2016jyx,Xie:2016fjj,Becattini:2017gcx} and transport-hybrid~\cite{Jiang:2016woz,Xia:2018tes,Wei:2018zfb,Wu:2019eyi} calculations 
predict a negative sign of the longitudinal component of the polarization vector in the first 
  quadrant of the ${\bf p}_T$ plane rotating counterclockwise from the reaction plane.
One such calculation is shown in the right panel of figure~\ref{fig:LongitudinalVersusPhi},
  while the corresponding measurement~\cite{Adam:2019srw} is on the left.
The magnitude of the effect is significantly larger in the model, but more strikingly, the sign
  of the predicted oscillation is opposite that seen in the data, reminiscent of the discrepancy in figure~\ref{fig:GlobalVersusPhi}.

Understanding and resolving the tension in figures~\ref{fig:GlobalVersusPhi} 
and~\ref{fig:LongitudinalVersusPhi} is among the most pressing open issues in this area.
This is further discussed in Section~\ref{issues}.

\section{OPEN ISSUES AND OUTLOOK}\label{issues}\label{sec:OpenQuestions}

Above, we have presented the theoretical framework (mostly hydrodynamics) in which to calculate 
the vorticity of the QGP; the theoretical connection between the vorticity and the polarization 
of hadrons emitted from the plasma, based on local thermodynamic equilibrium of hadrons and 
their generalized distribution function; and the measurements of this polarization with $\Lambda$ 
hyperons. Overall, the hydrodynamic and statistical equilibrium paradigm predicted first experimental 
observations of global polarization strikingly well.

However, qualitative discrepancies between theory and experiment may indicate that some fundamental 
feature of the dynamics itself (encoded hydro or transport) is misunderstood or unaccounted for.
Alternatively, we may misunderstand the interface (``Cooper-Frye" and thermal vorticity) between 
hydrodynamics or its coarse-grained approximation and the polarization observable. Clearly, the 
existing data demands more theoretical work and a report of the recent and ongoing work is the
subject of the next subsection.

On the other side, there are many important theoretical predictions which demand experimental tests. 
These will involve new detectors, future facilities, and new analysis techniques.

Finally, two topics deserve separate attention. One is the possibility that polarizations of 
$\Lambda$ and $\overline{\Lambda}$ are different. The other concerns the spin alignment of vector 
mesons.

\subsection{Local polarization}

The discrepancies between hydrodynamic calculations and the polarization pattern in momentum
space have been presented in subsection~\ref{sec:Longitudinal} and they have been the subject 
of investigations over the past year. 

The simplest explanation of them being an effect of secondary decays (see subsection~\ref{decays}) has 
been ruled out \cite{Xia:2019fjf},\cite{Becattini:2019ntv}; the secondary $\Lambda$'s have almost the same 
momentum dependence polarization as the primary, if all the primary are polarized according
to the hydrodynamic predictions. The other simple explanation is a polarization change in 
post-hadronization rescattering, which is not taken into account in simulation codes; however, 
this seems to be very unlikely, see discussion in subsection \ref{decays}, especially because
it should produce an amplification in some selected momentum regions. The available hadronic 
transport codes do not include the spin degrees of freedom mostly because the helicity-dependent 
scattering amplitudes are unknown and, even resorting to educated guesses, it is a formidable 
computational task to include them into Monte-Carlo codes. 

Within the hydrodynamic paradigm, there are more options yet to be explored. The first is 
concerned with the formula~\ref{basic}, which is first-order in thermal vorticity. Indeed, 
thermal vorticity is moderately smaller than 1 (see figure~\ref{vorthist}) and the exact formula 
at all orders is not known yet, so a sizeable role of higher order corrections cannot be
ruled out for the present.\\
Since polarization is steered by thermal vorticity, it is possible that the thermal vorticity 
field is different from the predictions obtained with the presently used initial hydrodynamic 
conditions, tuned to reproduce a set of observables in momentum space. Recently, ref.~\cite{Xie:2019jun} 
obtained the right sign of the longitudinal polarization at \ssn = 200 GeV with specific initial 
conditions \cite{Magas:2002ge}, while the same model predicts the "wrong" sign at lower energy 
\ssn = 8 GeV \cite{Xie:2016fjj}.\\
Another possibility is that spin dissipative corrections, analogous to viscous corrections for the
stress-energy tensor, which are not included in the local thermodynamic equilibrium assumption,
are sizeable. As has been mentioned in Section~\ref{theory}, the theory of dissipation and spin 
in hydrodynamic framework has recently drawn the attention of several authors; as yet, it is not 
clear whether such an approach includes relevant quantum terms and if it is pseudo-gauge dependent 
(see subsection \ref{spintensor}).\\ 
Furthermore, it has been considered that other kinds of vorticity, instead of thermal vorticity~\ref{thvort}, 
enter in the polarization definition. In ref.~\cite{Wu:2019eyi} it has been
shown that the right sign of the longitudinal polarization is retrieved if the thermal vorticity
is replaced by the a tensor proportional to the T-vorticity \cite{Becattini:2015ska}
whereas in ref.~\cite{Florkowski:2019voj} the agreement was restored by replacing the thermal 
vorticity with its double projection perpendicular to the velocity field. So far, these 
observations are not borne out by fundamental theoretical justifications.\\
Finally, it should be mentioned that in ref.~\cite{Liu:2019krs} the correct polarization patterns
have been obtained for the polarization of quarks within a chiral kinetic model; the question
remains on the effect of hadronization.

If all of the above ideas will fail to describe the data, two  scenarios may
be envisioned:
\begin{itemize}
    \item Spin does not locally equilibrate and it has to be described within a kinetic
    approach; c.f. section~\ref{kinetic}; 
    \item Spin equilibrates locally, but pseudo-gauge invariance is broken and one needs
    a spin potential to describe its hydrodynamic evolution, with six additional degrees of
    freedom and six additional hydrodynamic equations (see subsection~\ref{spintensor}).
\end{itemize}
Of course, both should be able to explain why the global polarization is in very good 
agreement with local equilibrium with thermal vorticity. Finally, we should always 
consider the possibility of a thus far unsuspected important ingredient.

\subsection{$\Lambda-\bar\Lambda$ splitting}
\label{sec:OpenQuestionSplitting}

As we have discussed, while the difference is statistically insignificant at any given energy, $P_{\overline{\Lambda},J}$ is systematically
  larger than $P_{\Lambda,J}$ at the lower collision energies where polarization itself is large.
A possible interpretation of such a splitting is the presence of a large electromagnetic field 
and one could use the observed difference to extract the value of the magnetic field in the 
rest frame of the particles, as discussed in subsection~\ref{sec:em}.

To first approximation, the $\hat{J}$-component of the vorticity is determined by the sum of $P_{\Lambda,J}$ and
  $P_{J,\overline{\Lambda}}$, and the magnetic field by their difference~\cite{Becattini:2016gvu}.
However, feed-down corrections can be important, and should be accounted for~\cite{Becattini:2016gvu}.
For example, in the absence of feed-down, a finite $B$-field would produce
  $P_{\overline{\Lambda},J}>P_{\Lambda,J}$, and $B=0$ would result
  in no difference in the polarizations.
However, if $B=0$, feed-down effects at low collision energies (where there are significant chemical potentials at freezeout)
  can generate a "splitting" with opposite sign, i.e.  $P_{\overline{\Lambda},J}<P_{\Lambda,J}$~\cite{Becattini:2016gvu}.
Applying formula~\ref{basic} (with the substitution~\ref{emfield}) to the data in figure~\ref{fig:RootsDependence}, and
  accounting for feeddown effects~\cite{Becattini:2016gvu}, results~\cite{LisaQM2019} in an estimate of $B=(6\pm6)\times10^{13}$~T when averaging
  over results from $10~{\rm GeV}<\sqrt{s_{NN}}<40{\rm ~GeV}$.
Such an average is hardly justified, but it nevertheless provides a valuable estimate of the magnitudes of the magnetic field-- and the
  measurement uncertainty-- that may be associated with the data.
In the equilibrium paradigm, this is the magnetic field at freezeout.
Theoretically, fields of this magnitude are present in the first instants of a heavy ion collision.
While they may decay well before freeze-out, a highly conductive QGP itself can
  significantly extend the lifetime of the initially large field~\cite{McLerran:2013hla}
  and vorticity certainly helps in this respect~\cite{Guo:2019mgh}. At low $\sqrt{s_{NN}}$, field lifetimes 
  may be longer~\cite{Skokov:2009qp} and QGP evolution time shorter. 
  Relativistic magneto-hydrodynamics it the standard tool to study the evolution of the
  electro-magnetic field in a plasma and there have been major advances recently
  \cite{Inghirami:2016iru,Inghirami:2019mkc}. 
Neglecting feed-down corrections, the similarity of $P_{\Lambda,J}$ and $P_{\overline{\Lambda,J}}$ places~\cite{Muller:2018ibh}
 an upper limit on the magnetic field at freezeout of about $10^{12}$~T at top RHIC energy.

Transport calculations may provide more insight, with respect to local thermodynamic equilibrium.
Simplified calculations estimate that the expected field could be on order $10^{12}-10^{13}$~T, 
and the energy dependence of the splitting would resemble that seen in the data~\cite{Guo:2019mgh}.
A more sophisticated calculation~\cite{Guo:2019joy} with partonic transport argues that the difference between $P_{\Lambda,J}$ and $P_{\overline{\Lambda},J}$ may be reasonably attributed to the accumulated 
effect of an evolving magnetic field; interestingly, in the absence of a magnetic field, $P_{\overline{\Lambda},J}<P_{\Lambda,J}$.
  
  
A firm statement on the existence of a long-lived (several fm/c) magnetic field on the scale of $10^{13}$~T would
  have tremendous implications for the Chiral Magnetic Effect~\cite{Fukushima:2008xe}.
However, several effects have been postulated, which may complicate the interpretation of the splitting.
Especially at low collision energies where baryon stopping is significant, $\Lambda$ and $\overline{\Lambda}$ may
  originate from different regions within the fireball~\cite{Vitiuk:2019rfv}, their final polarizations thus
  reflecting differently-weighted averages over vorticity.
Han {\it et al} argue that the quark-antiquark vector potential in the presence of the net quark flux at these energies
  may generate a splitting largely indistinguishable that expected from a magnetic field.
The vector meson field may play an equivalent role in the hadronic sector; however, existing calculations~\cite{Csernai:2018yok,Xie:2019npz}
  reproduce the splitting only by adjusting by hand the unknown sign, magnitude and energy dependence of the effect.
  
In principle, disentangling these effects could require
  a full three-dimensional magnetohydrodynamic calculation which includes appropriate vector potentials, conserved
   nontrivial baryon currents and QGP conductivity, potentially followed by a hadronic cascade
   with spin-transfer collisional dynamics.
Hopefully, however, sophisticated but achievable calculations, in conjunction with targeted measurements,
  can lead to reasonable estimates of the individual contributions of these important effects.

\subsection{Alignment}
\label{sec:align}

In peripheral collisions the anisotropy generated by the collective angular momentum implies 
that all particles with spin can, in principle, have a non-vanishing polarization. 
Particularly, for vector mesons, this implies a non-vanishing alignment, as discussed 
in subsection~\ref{sec:MeasuringPolarization}. At local thermodynamic equilibrium, the 
equation \ref{spindensmat2} predicts an alignment which is quadratic in the thermal vorticity 
\cite{Becattini:2007nd}. Since thermal vorticity is less than 1 throughout (see figure 
\ref{vorthist}), and at freezeout it is on order 0.02, the expected resulting alignment is tiny. 

In fact, preliminary results on the alignment have been reported for two vector mesons, $K^*$ and $\phi$ at RHIC~\cite{Singha:2020qns}
  and LHC~\cite{KunduQM2019}.
In all cases, $\Theta_{00}$ (c.f. equation~\ref{eq:SpinAlignment}) is considerably different than $\tfrac{1}{3}$.
For $\phi$ at LHC and $K^*$ at both colliders would imply a vorticity at least two orders of magnitude higher than calculations or
  expectations from the hyperon measurements.
More surprisingly, $\Theta_{00}$ for the $\phi$ mesons at RHIC is {\it greater} than $\tfrac{1}{3}$~\cite{Singha:2020qns},
  something that
  cannot be understood in hydrodynamic or recombination model~\cite{Liang:2004xn}.
This observation may require fundamentally new physics mechanisms~\cite{Sheng:2019kmk} for alignment
  that apply at RHIC but not at LHC.
Altogether, the situation with these preliminary spin alignment measurements is 
  not sufficiently well understood to discuss in a review.

\subsection{Future measurements}
\label{sec:FutureExperiment}

As we have discussed, the first few positive observations of hyperon polarization at RHIC 
have generated tremendous theoretical activity. Much of this work has focused on the degree 
to which models can reproduce the measurements, but a growing body of work points to the 
ways in which new measurements can strictly test our understanding of QGP dynamics and may 
provide enhanced sensitivity to important physics.

\noindent{\bf Lower energy collisions}\\
As seen in figure~\ref{fig:RootsDependence}, both the observed and predicted polarization signals
  rise as the collision energy is reduced.
Exploring this trend for even lower energies may touch on several important questions:
does a hydrodynamic description of the system break down at lower energy density?
What are the effects of increased viscosity~\cite{Wang:2013xtp,Karpenko:2016jyx,Karpenko:2018erl,Montenegro:2018bcf}?
Can spin equilibrate rapidly enough to justify a local thermodynamic equilibrium approximation 
-- and if so, is this due to hadronic mechanisms or to the QCD phase transition?
Some hydrodynamic models tuned for low energies predict an uninterrupted continued 
rise~\cite{Csernai:2014ywa,Xie:2016fjj}, regardless of equation of   
  state~\cite{Kolomeitsev:2018svb,Ivanov:2019ern}, though initial state and thermalization assumptions
  may affect this behavior at the lowest energies,
  producing a non-monotonic behavior~\cite{Deng:2020ygd}.

In section~\ref{sec:GlobalData}, we remarked on the possible tension
  between $\Lambda$ and $\overline{\Lambda}$ directed flow and polarization at low 
  collision energy; how tightly coupled are the large-scale and smaller-scale rotational 
  structures in the flow fields probed by these particles?
It has been suggested~\cite{Ivanov:2014ioa,Nara:2019qfd} that
  the diverging behavior of baryon and antibaryon directed flow~\cite{Adamczyk:2014ipa,Adamczyk:2017nxg} 
  signals a phase transition in the equation of state.
Alternatively, it may arise from a convolution of baryon stopping and 
  quark coalescence~\cite{Dunlop:2011cf,Adamczyk:2017nxg}.

Again regarding $\Lambda$ and $\overline{\Lambda}$, it has been suggested~\cite{Vitiuk:2019rfv} that
  differences in polarization are dominated by differences in the phase space from 
   which these particles arise; such differences are largest at low collision energy.
Testing this hypothesis and that discussed in the previous paragraph of course will require 
measurement of local polarization, that is as a function of momentum.

Addressing these questions will require new measurements at $\sqrt{s_{NN}}\lesssim10$~GeV, with
  good tracking, event plane resolution, and high statistics, especially
  given plummeting $\overline{\Lambda}$ yields.
These will be pursued at the future NICA~\cite{Kekelidze:2017tgp,Golovatyuk:2019rkb} and FAIR~\cite{Schmidt:2014lva} 
  facilities, as well as the STAR/RHIC fixed-target program~\cite{Meehan:2017cum} and
  the HADES/GSI experiment~\cite{Agakishiev:2009am}.
  
\noindent{\bf Measurements at forward rapidity}\\
Thus far, polarization has been measured at midrapidity, to focus on the hottest part of the
  QGP fireball.
However, calculations with a variety of models suggest a vortical structure that evolves with
  rapidity.
  
A geometric calculation~\cite{Betz:2007kg} based on the BGK model of 
  hadron production~\cite{Brodsky:1977de} and boost-invariance suggests that vorticity
  will increase with rapidity, and speculates that a rapid change in the evolution could
  signal a phase change at some critical density.
A similar, more recent calculation~\cite{Liang:2019pst} finds that the rapidity dependence
  of vorticity itself depends on $\sqrt{s_{NN}}$ at RHIC energies, and that it is sensitive
  to important physical parameters of the model itself.
Numerical calculations with transport-hybrid codes~\cite{Jiang:2016woz,Deng:2016gyh,Wu:2019eyi}
  also indicate a forward migration of vorticity, especially as the collision energy 
  increases~\cite{Deng:2016gyh}. Finally, hydrodynamic models predict much higher vorticity 
  in the beam fragmentation region at both NICA~\cite{Baznat:2015eca,Ivanov:2017dff} and 
  RHIC~\cite{Ivanov:2018eej,Ivanov:2019ern} energies.
  
Exploring vorticity away from midrapidity in fixed-target experiments (discussed above) is
  relatively straightforward.
At collider energies, the STAR forward upgrade will provide coverage and tracking over a physically
  important region~\cite{Liang:2019pst}.
If the event plane can be reconstructed, the LHCb experiment~\cite{Alves:2008zz}
  could be used to explore the rapidity evolution at the highest energies.

\noindent{\bf Other polarization projections}\\
Referring to equation~\ref{eq:HyperonDecay2}, experiments have reported polarization projections along 
  $\hat{n}=\hat{J}$ and $\hat{n}=\hat{p}_{\rm beam}$.
The geometry of the collision itself suggests other natural directions.

For particles emitted at forward rapidity, symmetry permits an average polarization projection
  along $\hat{n}=\hat{b}$.
In fact, "vortex rings" or "cyclones" are predicted~\cite{Xia:2018tes,Ivanov:2018eej} at forward rapidity at
  RHIC~\cite{Xia:2018tes} or NICA~\cite{Baznat:2015eca,Ivanov:2018eej} energies, as well as
  at midrapidity in non-symmetric systems~\cite{Voloshin:2017kqp}.
In this case, $\hat{n}\parallel\vec{p}_{\Lambda}\times\hat{z}$.

One of the first model studies of vorticity in heavy ion collisions predicted similar ring-like
  structures relative to jets.
High-momentum partons formed in the initial stages of the heavy ion collision
  lose energy in the QGP fireball~\cite{Majumder:2010qh} and can locally perturb the
  flow field~\cite{Tachibana:2020atn}.
This may produce a cone or ring of vortical structure locally perpendicular to the direction of the
  deposited momentum~\cite{Betz:2007kg}, $\hat{n}=\hat{p}_{\rm dep}\times\hat{p}_{H}$,
  where the hyperon $H$ has acquired an outward velocity from the radial flow~\cite{Heinz:2013th} of the QGP.
  
Finally, the QGP depicted in figure~\ref{fig:CollisionSketch} is likely to be characterized by
  turbulence~\cite{Florchinger:2011qf,Csernai:2011qq,Csernai:2014nva}, in which the vorticity of
  a fluid cell is not correlated with a global event characteristic or symmetry-breaking direction.
However, the assumption is that the polarizations of all particles emitted from a cell are aligned
  with the vorticity of that cell, and flow-induced space-momentum correlations~\cite{Heinz:2013th}
  cause particles from the same cell to be emitted in the same direction.
Hence, if experimental complications can be overcome~\cite{LisaSpringer:2020},
  spin-spin correlations as a function of relative momentum (or angle) are a promising way
  to probe the turbulent vortical substructure of the QGP~\cite{Pang:2016igs}.

\section{SUMMARY AND OUTLOOK} 

Polarization has opened an exciting new direction in relativistic heavy ion physics; one of 
the increasingly rare truly new developments in this rather mature field.
Its measurement has definitely proved that a new degree of freedom other than momentum is now
available to probe the QGP formation and dynamics. In the hydrodynamic model, unlike particle
momentum, polarization is primarily sensitive to the gradients of the hydro-thermal fields, 
and this appears to be a unique feature among the known observables. Moreover, polarization 
can help to constrain the electro-magnetic field, which would be incredibly valuable for the 
search of Chiral Magnetic Effect \cite{Fukushima:2008xe}.
The hydrodynamic model predicts, and the measurements have shown, that polarization increases at
low energy, and it will be further explored in future low-energy heavy 
ion programs. 
At RHIC and LHC energies, flow substructure is already being probed in unprecedented detail, 
  presenting theory with new and as yet unsolved challenges.
Directions for future studies at these energies were discussed.

There are several pressing issues to be solved which require considerable advances in
theory and phenomenology. Indeed, at this time, after having played the leading role, theory 
appears to have been surpassed by the experiments
which have proved to be able to measure polarization as a function of many relevant variables
in relativistic heavy ion collisions: azimuthal angle, rapidity, centrality, etc.
In the near future, more measurements will be available which will help to constrain or disprove
theoretical models and assumptions; polarization of different species (e.g. $\Sigma^0$ and $\Xi^-$),
spin-spin correlations \cite{Pang:2016igs}; measurement of polarization in different colliding 
systems~\cite{Shi:2017wpk}.
On the theory side, as has been mentioned, one expects improved formulae including more terms
and corrections to the equation~\ref{basic}, the inclusion of dissipative effects and the application of 
alternative methods such as kinetic theory as well as the development of a hydrodynamic with
spin potential. Equally important is a major advance in phenomenology and numerical computation, 
with the inclusion of hadronic rescattering effects and the systematic study of polarization 
dependence on the initial conditions. 

Since its experimental discovery a few years ago, there has been tremendous progress in the
  study of polarization in heavy ion collisions. Yet, at this early stage, the potential 
  of this new tool is still to be explored. 
It may well be that this direction of research yields new insights and major results
in the study of the QCD matter with nuclear collisions.

\section*{DISCLOSURE STATEMENT}
The authors are not aware of any affiliations, memberships, funding, or financial holdings that
might be perceived as affecting the objectivity of this review.

\section*{ACKNOWLEDGMENTS}

We are greatly indebted to Gabriele Inghirami for his invaluable help in making some
of the figures of this article.
F. B. was partly supported by the INFN project SIM.
M.A.L. supported by the U.S. Department of Energy.

%

\end{document}